\documentclass[pra,twocolumn,groupedaddress]{revtex4-1}

\usepackage{graphicx}
\usepackage{dcolumn}
\usepackage{bm}
\usepackage[english]{babel}
\usepackage{epsfig}
\usepackage{mathtext}
\usepackage{epstopdf}
\usepackage[mathlines]{lineno}
\usepackage{amsmath,amssymb,graphicx,color}
\usepackage{color}
\usepackage{array}
\newcolumntype{P}[1]{>{\centering\arraybackslash}m{#1}}
\graphicspath{{test/}}

\begin{document}

\title{Enhancement of Terahertz Photoconductive Antennas and Photomixers Operation by Optical Nanoantennas}

\author{Sergey~I.~Lepeshov,$^{1}$ Andrei~Gorodetsky,$^{1,2}$ Alexander~E.~Krasnok,$^{1}$ Edik~U.~Rafailov,$^{2}$ and Pavel~A.~Belov$^{1}$}
\address{
$^{1}$ITMO University, St.~Petersburg 197101, Russia\\
$^{2}$Aston Institute of Photonic Technologies, Aston University, B4 7ET, Birmingham, UK}

\begin{abstract} Photoconductive antennas and photomixers are very promising sources of terahertz radiation that is widely used for spectroscopy, characterisation and imaging of biological objects, deep space studies, scanning of surfaces and detection of potentially hazardous substances. These antennas are compact and allow generation of both ultrabroadband pulse and tunable continuous wave terahertz signal at room temperatures, without a need of high-power optical sources. However, such antennas have relatively low energy conversion efficiency of femtosecond laser pulses or two close pump wavelengths (photomixer) into the pulsed and continuous terahertz radiation, correspondingly. Recently, an approach to solving this problem has been proposed, that involves known methods of nanophotonics applied to terahertz photoconductive antennas and photomixers. This approach comprises the use of optical nanoantennas for enhancing the efficiency of pump laser radiation absorption in the antenna gap, reducing the lifetime of photoexcited carriers, and improving antenna thermal efficiency. This Review is intended to systematise all the results obtained by researchers in this promising field of hybrid optical-to-terahertz photoconductive antennas and photomixers.\end{abstract}

\maketitle

\tableofcontents
\begin{figure*}
\centering
\includegraphics[width=0.99\textwidth]{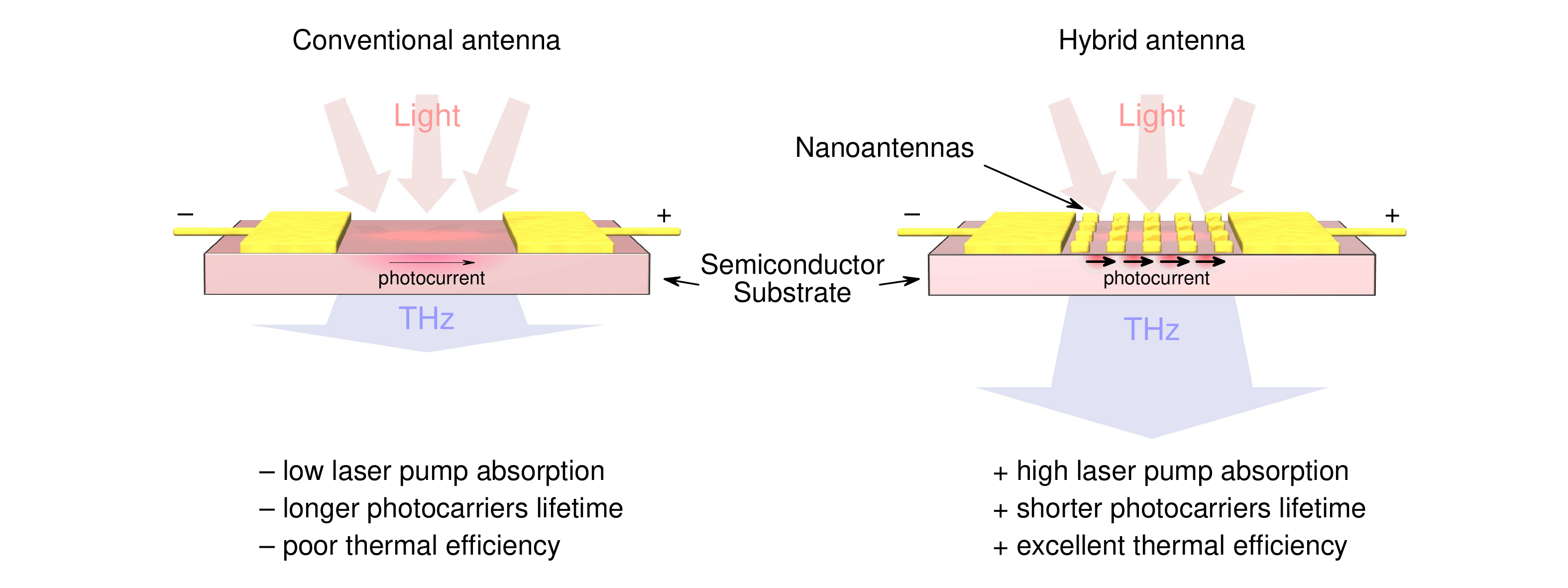}
\caption{Illustration of the hybrid antenna advantages over conventional one.}
\label{0}
\end{figure*}

\section*{Introduction}

Terahertz (THz) range of electromagnetic waves tightly seated between microwave and optical regions, is of great interest, primarily due to the fact that in this band of the electromagnetic spectrum reside frequencies of some elementary excitations in semiconductors and dielectrics~\cite{Ferguson2002a, Schmuttenmaer2004, Tonouchi2007, rev1, semicon1, semicon2, dielec1, supercon1, supercon2, supercon3}, as well as rotational and vibrational spectra of complex, including biological, molecules~\cite{Zeitler2007, poly1, poly2, poly3, poly4, protein1, protein2, protein4, protein5}. As a result, THz waves have tremendous application capacity in areas ranging from the detection of dangerous and illicit substances~\cite{Federici2005, secure1, secure2, secure3, rev1, Zeitler2007} to diagnosis and treatment of diseases in medicine~\cite{Davies2002, protein4, protein5, med1, med5, Ashworth2009, Wallace2006,Shiraga2014, Rong2015}. Moreover, this frequency band is of a special importance for spectroscopy in astrophysics~\cite{astr1, astr2, astr3, astr4, astr5, astr6} for the reason that many astrophysical objects emit in the THz range of the spectrum. In addition, THz technology is promising for wireless communication systems as a replacement for transmitters and receivers operating in high-frequency and microwave ranges, because it can potentially increase the speed of data transfer by hundreds of times~\cite{rev5, com_ap2, com_ap3, com_ap4, com_ap5, com_ap6}. Although, there are hurdles, like strong absorption of THz waves by metals and water, on the way to the replacement of mega- and gigahertz frequency devices~\cite{water_loss1, water_loss2, water_loss3}, a significant progress has already been made~\cite{Tonouchi2007, com_ap1, com_ap7, com_ap8}.

Until relatively recently, the generation of coherent THz waves (in other words, submillimeter radiation) was an extremely difficult task. However, research in the semiconductors~\cite{Vodopyanov2008, Saeedkia2008, semicon1, semicon2, semicon_res1, semicon_res2, Leyman2016}, and the interaction of short (compared with all relaxation times) optical pulses with semiconductor and nonlinear materials (\textit{photoconductivity effect}~\cite{photocon1,photocon2}) stimulated the development of compact low-power THz radiation transmitters. One of the most simple but not the most efficient way to obtain the submillimeter waves is by acceleration of movement of the separated charge carriers in the surface field of a semiconductor under ultrashort light pulses~\cite{photocon1, semicon1}. One of the physical mechanisms that drives the carriers inside a semiconductor is \textit{photo-Dember effect}. Photo-Dember effect is the formation of the electric field (and consequently a charge dipole) in the illuminated semiconductor due to the difference of electrons and holes diffusion velocities~\cite{dember} under the strong absorption of optical excitation, which leads to an effective charge separation in the direction perpendicular to the semiconductor surface~\cite{Johnston_2001}. Another possible way to generate THz -- \textit{optical rectification} -- works when media obtains non-linear polarisation under intense optical radiation, and results in polarisation repeating the shape of the optical pulse envelope~\cite{rectif1, rectif2, rectif3}. Moreover, apart from these well-established methods, novel approaches of photonics to THz generation that employ artificial electromagnetic medium (\textit{metamaterials and metasurfaces}) have been recently proposed~\cite{McCutcheon2009, Docherty2012, Luo2014, Barnes_2014, Evtikhiev_2014, Zhang_2015, Seifert2015, Nakajima_2016, Yanagi_2016}.

Nowadays, the most common method of low power THz generation involves the use of semiconductor structures surface conductivity~\cite{electrodes1, electrodes2}. Two or more conductive electrodes spaced by a certain gap are deposited onto semiconductor surface (Figure~\ref{0}, left side). The electrodes are biased by the external voltage of several $\rm V$. Such structure is referred to as \textit{THz photoconductive antenna} if pumped by femtosecond optical pulses, or \textit{THz photomixer} in the case of continuous (CW) pump by two lasers operating at close wavelengths with a difference frequency of these wavelengths being in the THz spectral range. When exciting the gap between the electrodes with a femtosecond laser, the concentration of charge carriers increases sharply for a short period of time. At this point, THz pulse generation occurs. The duration of this pulse and its spectrum are determined mainly by the carriers lifetime in the semiconductor. In the case of CW pumping, carrier concentration changes with difference frequency of pump wavelengths, and CW THz signal is emitted. The efficiency of such THz sources is strongly limited by the amount of the optical energy of laser radiation absorbed in the gap of photoconductive antenna. Usually this amount is rather small, taking into account the high values of the semiconductor refractive index at optical frequencies, leading to a high reflection coefficient. Moreover, the effectiveness of traditional photoconductive antennas is limited by low drift velocities of the photoinduced charge carriers in semiconductor substrates~\cite{3d42, 3d43, carrier1}.

It has been shown recently, that the efficiency of a THz pulse generation can be significantly increased by placing the so-called \textit{optical nanoantennas}~\cite{Quate_2003, 26, optic2, optic3, optic5, optic6, optic1, Argyropoulos_2015, Hulst_2011, Krasnok_SC_2015} into the photoconductive gap (Figure~\ref{0}, right side)~\cite{mona, geom2, geom3, electrodes1, electrodes2, resist, grate1, gex1, 3d75}. As a result of such \textit{integration}, one gets an antenna comprising a comprehensive photoconductive THz antenna, and an array of optical nanoantennas (see Figure~\ref{0}). Such antenna can be called a \textit{hybrid THz-optical photoconductive antenna}. These antennas are the result of application of known methods of \textit{nanophotonics} to the area of efficient sources of THz radiation. The efficiency of the entire hybrid antenna is determined by the efficiencies of THz photoconductive antenna and an optical nanoantenna array, and is usually higher by at least one order of magnitude. Moreover, optical nanoantennas embedded in a THz photoconductive antenna improve a thermal stability of the latter, due to the large thermal conductivity of metals which they are made of. As for detectors, the use of hybrid photoconductive THz detectors, with nanoparticles in the photoconductive gap, allowed the enhancement of the near field THz imaging resolution down to $\lambda/150$~\cite{Mitrofanov2015a}. Such hybrid THz-optical antennas will be discussed in this Review.

It should be noted that the present Review is the first attempt to collect all currently available literature about hybrid THz-optical antennas. The aim of this Review is to demonstrate recent achievements in this field. We will not focus on the current research in the field of conventional photoconductive antennas and photomixers, but will briefly describe their main operating principles. A great number of excellent reviews have been written on this subject, among them we strongly recommend the following~\cite{rev1, rev2, rev3, rev5, Reimann2007, Vodopyanov2008}. Our Review is primarily addressed to specialists working with THz radiation and its compact sources, but is written in such a way that both engineers and physicists working in related areas, can study the methods of THz generation using semiconductors and then obtain an information about the latest achievements in the field of hybrid photoconductive antennas.

The Review is structured as follows. In Part~\ref{first} the most general principles of THz generation in semiconductor sources pumped with femtosecond optical pulses and double-wavelength CW laser radiation are described: in Section~\ref{first1} the mechanisms of pulsed THz radiation generation in semiconductors are considered in more detail; Section~\ref{first2} is completely dedicated to the principles of THz photoconductive antennas; Section~\ref{first3} describes the principles of THz photomixers; Section~\ref{first4} contains the information about the main characteristics of such antennas. Part~\ref{second} is entirely devoted to the review of hybrid THz-optical photoconductive antennas: Section~\ref{second01} outlines the general principles of operation of optical nanoantennas and the prospects they offer; Section~\ref{second3} describes the plasmon monopoles as an effective way to enhance the conversion of light into current; Section~\ref{second0} contains a detailed description of metal dipole nanoantennas, their types and features of their operation in hybrid antennas; Section~\ref{second1} is devoted to plasmon gratings as absorption amplifiers in the gaps of the photoconductive antennas; Section~\ref{second2} reviews the works on three-dimensional plasmonic gratings. Part~\ref{third} is the discussion and outlook, which summarises all the results described in detail in Part~\ref{second}, compares all the approaches to the implementation of the hybrid antennas, and offers further perspectives of their development such as \textit{all-dielectric nanoantennas} and the search for new materials.

\section{THz photoconductive antennas and photomixers}\label{first}

\subsection{The mechanism of pulsed THz radiation in semiconductors}\label{first1}

Currently, there are many different approaches to THz generation, in a vast variety of layouts and involving various materials. However, for practical applications, a method involving the excitation of non-equilibrium charge carriers in the surface layer of semiconductor structures by ultrashort optical pulses is the most widely used now. Figure~\ref{1}(a) shows the simplest layout for sub-millimetre waves generation in semiconductor based photoconductive antenna. \textit{Femtosecond laser} (usually Ti:Sapphire or fibre~\cite{rev1}) serves as a source of ultrashort light pulses with a duration of several tens or hundreds of femtoseconds. The shape of the optical pulse envelope is important, as it induces the nonlinear polarization of the same shape therein (optical rectification), which is capable to increase the output power of THz radiation~\cite{rectif1,rectif2}. \textit{Silicon lens}, shown in Figure~\ref{1}(a), serves as a collecting lens for THz radiation~\cite{lens1, lens2} and at the same time as a heatsink to transfer an excess of heat from the substrate, and as an anti-reflective element, since the refractive index of silicon in THz is equal to $n_{_{\rm Si}}=3.4$~\cite{Grischkowsky1990}, which is about the same as one of high-resistive GaAs ($n_{_{\rm GaAs}}\approx3.6$)~\cite{Grischkowsky1990}.

Let us consider the model of THz generation in a semiconductor. The dependence of the output power $P_{_{\rm{THz}}}$ on the frequency $\omega_{_{\rm THz}}$ is as follows~\cite{optim}:

\begin{equation}
P_{_{\rm THz}}\left(\omega_{_{\rm THz}}\right) \backsim \frac{1}{1+(\omega_{_{\rm THz}}\tau_{_{\rm eff}})^2},
\label{eqP}
\end{equation}
where $\tau_{_{\rm eff}}$ is the effective lifetime of photocarriers, equal to the lifetime of nonequilibrium charge carriers in the semiconductor. From the equation (\ref{eqP}), it is clearly seen that the shorter the lifetime $\tau_{_{\rm eff}}$ is, the greater power $P_{_{\rm THz}}$ can be obtained. Therefore, gallium arsenide grown by molecular beam epitaxy on a low-temperature (LT) substrate (LT-GaAs) is usually chosen as a photoconductive material, as this growth method allows to obtain a material with more defects and hence with a shorter lifetimes of nonequilibrium charge carriers ($\le$~200~fs~\cite{optim}). When irradiated by femtosecond laser pulses with photon energies greater than the bandgap, the concentration of nonequilibrium charge carriers in LT-GaAs rapidly increases, and in the surface layer electron-hole pairs are formed. In the semiconductor model all electrons are set to have the same effective mass, as they have a low energy because of the wide band of gallium arsenide. The model is not applicable to narrow-gap semiconductors like indium arsenide (InAs). In the case of InAs, it is necessary to consider ensembles of equilibrium and photoexcited electrons interacting through self-consistent field~\cite{magn2}, which greatly complicates the numerical modeling and description of processes.

Due to the localisation of charges at surface states in the surface layer, so-called \textit{built-in field} appears. The light induced electrons and holes are accelerated and separated by this field. The resulting uneven distribution of charge is considered as a dipole with dipole moment equal to the sum of momenta of all induced charges in the volume. Dipole oscillations occur until the equilibrium is established again. During the relaxation, electrons emit energy in the form of electromagnetic waves in the THz range, and the relaxation time determines both the spectrum of this radiation and its power. Built-in surface field plays a very important role in this process. The stronger this field is, the more intensive THz signal is emitted. Therefore, metal electrodes are often applied to the semiconductor surface and biased to create stronger electric field. The mechanism of THz generation enhancement by such structures will be described in detail in the next section of this Review. Due to the fact that the generation takes place near the air-semiconductor interface, a larger portion of the THz radiation is emitted into the GaAs substrate. The reason for this is the greater optical density of the GaAs in comparison with air. Therefore, highly doped semiconductor materials are not suitable for efficient photoconductive antennas due to significant absorption in the THz region of the spectrum~\cite{Grischkowsky1990}. On the other hand, the doping of the sample increases the concentration of photoinduced electron-hole pairs, which, ultimately, increases the current and hence THz radiation power. A reasonable compromise between these factors (i.e. higher absorption vs higher current) will substantially improve the efficiency of generation.

Another effective method of increasing the efficiency of generation is the application of an external magnetic field. It has been shown, that a semiconductor placed in a constant magnetic field of 1~T generates THz radiation with efficiency up to 20 times larger than without magnetic field~\cite{magn4}. Magnetic field optimal direction is the one perpendicular to the semiconductor surface, and THz pulse amplitude is proportional to the magnetic field squared. This effect is explained by the appearance of the \textit{Hall component of the current}, excited carriers being driven by external or surface field~\cite{magn1,magn2,magn3}. The analysis of temperature dependence of submillimeter radiation power shows that with decreasing temperature the power goes up. With the GaAs temperature decreasing from 280~K down to 80~K, the radiation energy is increased by 3.4 times~\cite{Zhang1992}.

\subsection{Pulse THz generation in photoconductive antennas} \label{first2}

A way of THz generation in a semiconductor surface by the accelerated motion of nonequilibrium charge carriers driven by semiconductor surface field and induced by femtosecond laser pulses, has, however, low optical-to-terahertz conversion efficiency. Therefore, metal electrodes (usually a nickel-gold-platinum alloy) are deposited onto the substrate surface and are biased to create high electric field in in the gap between them, where photocarriers are generated, and thus enhance the generation of the THz waves~\cite{Auston1984}. {Such structures together with a photoconductive semiconductor substrate are referred to as THz photoconductive antennas. An example and principle of operation of such antenna is shown in Figure~\ref{1}(a).
\begin{figure}[!t]
\includegraphics[width=0.5\textwidth]{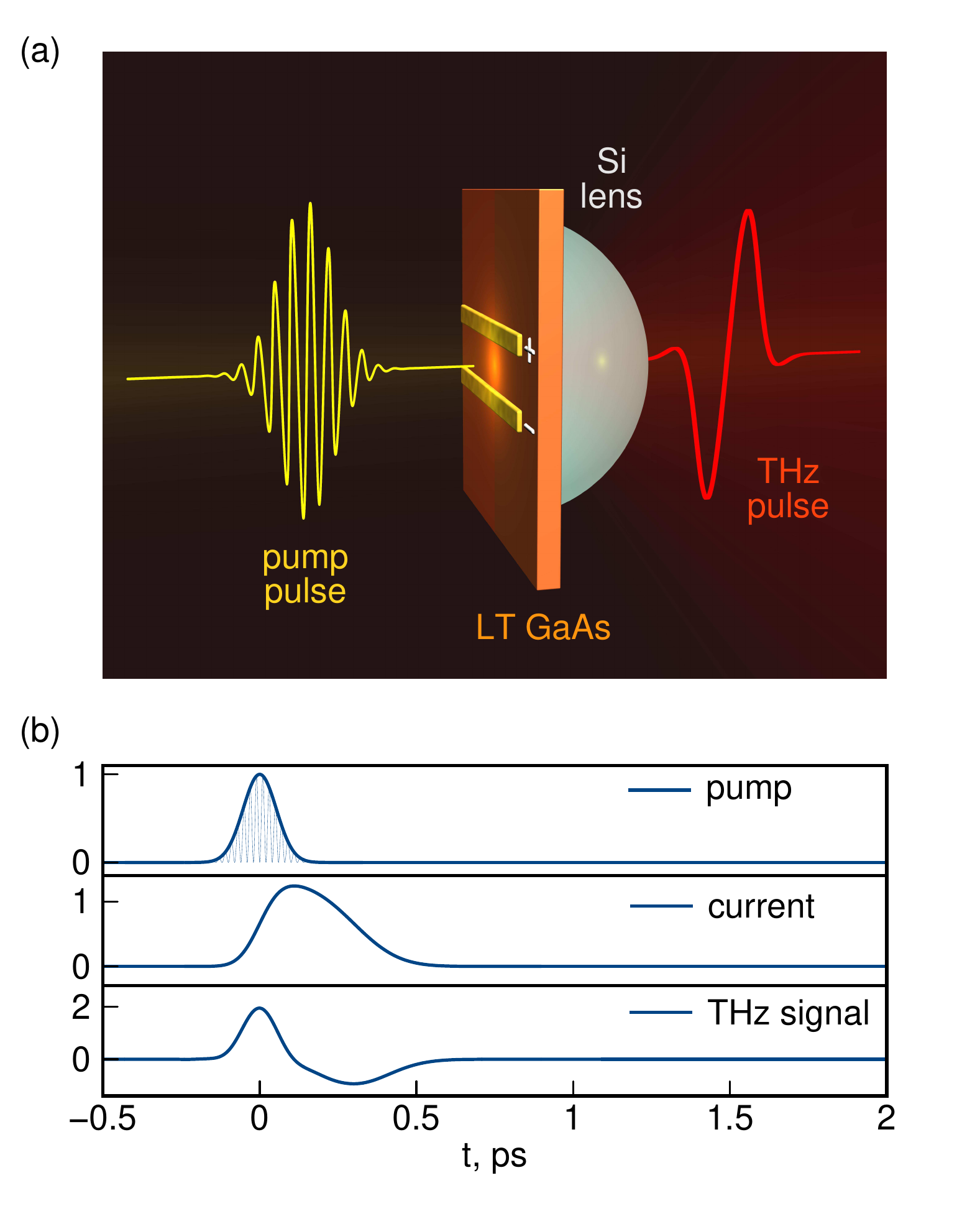}
\caption{Schematic view of a setup used to generate pulsed THz radiation by photoconductive LT-GaAs based antenna (a) and time evolution of pump radiation envelope, antenna photocurrent and generated THz signal (b).}
\label{1}
\end{figure}

The mechanism of THz generation in this case is similar to that discussed in the preceding section. Under the ultrashort laser pulse, electron-hole pairs are formed in the surface layer. The conductivity of the semiconductor as a consequence greatly increases proportional to electrons and holes concentration: $\sigma = qn_{\rm e}\mu_{\rm e} + qn_{\rm p}\mu_{\rm p}$, where {\em q} is the elementary charge, \begin{math}n_{\rm e}\end{math} and \begin{math}n_{\rm p}\end{math} are the concentrations of electrons and holes, \begin{math}\mu_{\rm e}\end{math} and \begin{math}\mu_{\rm p}\end{math} are their mobilities. The increase of conductivity results in a giant current pulse of photoinduced charges, electrons and holes being accelerated in opposite directions by the external electric field in the gap. The acceleration of charge carriers by the external field \begin{math}E_{\rm bias}\end{math} in this case can be described by the differential equation:
\begin{equation}
\frac{dv_{\rm e,p}}{dt} = -\frac{v_{\rm e,p}}{\tau_{\rm eff}} + \frac{q E}{m_{\rm e,p}},
\label{eq6}
\end{equation}
where \begin{math}v_{\rm e,p}\end{math} is the drift velocity of electrons and holes, \begin{math}m_{\rm e,p}\end{math} are their masses, \begin{math}\tau_{\rm eff}\end{math} is approximately 30~fs in low-temperature GaAs, \begin{math}E\end{math} is a local field, which is related to \begin{math}E_{\rm bias}\end{math} with the following ratio: $E = E_ {\rm bias} - \frac{P}{\alpha\varepsilon_{\rm 0}}$, where \begin{math}\alpha\end{math} is the static dielectric susceptibility of semiconductor (equal to 3 for the LT-GaAs), \begin{math}\varepsilon_{\rm 0}\end{math} is the vacuum permittivity, \begin{math}P\end{math} is the polarization caused by the separation of electrons and holes. The time evolution of polarization can be described by the following inhomogeneous differential equation:
\begin{equation}\frac{dP}{dt} + \frac{P}{\tau_{\rm rec}} = J(t)\label{eq9},\end{equation}
where \begin{math}\tau_{\rm rec}\end{math} is the recombination time of holes and electrons, and $J$ is the density of currents on the surface of the semiconductor defined by: $J(t) = qn_{\rm e}v_{\rm e} + qn_{\rm p}v_{\rm p}$. From expressions (\ref{eq6}-\ref{eq9}), using Maxwell's equations, the authors of the article~\cite{formul1} derived a formula for the intensity of the radiated THz field at a much larger than THz wavelength \begin{math}\lambda\end{math} distance z from the source:
\begin{equation}
E(z,t) = -\frac{A}{4\pi\varepsilon_{\rm 0}c^2z}\cdot\frac{dJ}{dt} = -\frac{A}{4\pi\varepsilon_{\rm 0}c^2z}\cdot(qv\frac{dn}{dt} + qn\frac{dv}{dt})
\label{eq11},
\end{equation}
where $n$ is the total concentration of carriers, $v$ is their velocity, $A$ is the area of radiating surface, equal to the area of the gap between the electrodes, and \begin{math}c\end{math} is the speed of light in vacuum. Thus, the external field increases the energy of charge carriers, and hence the output power of THz radiation, not influencing, however, the spectral characteristics, as it is determined by the lifetime of electrons and holes. Pump pulse envelope, photocurrent in the surface layer and resulting THz field examples obtained with this model are shown in Figure~\ref{1}(b).
\begin{figure}[!t]
\includegraphics[width=0.5\textwidth]{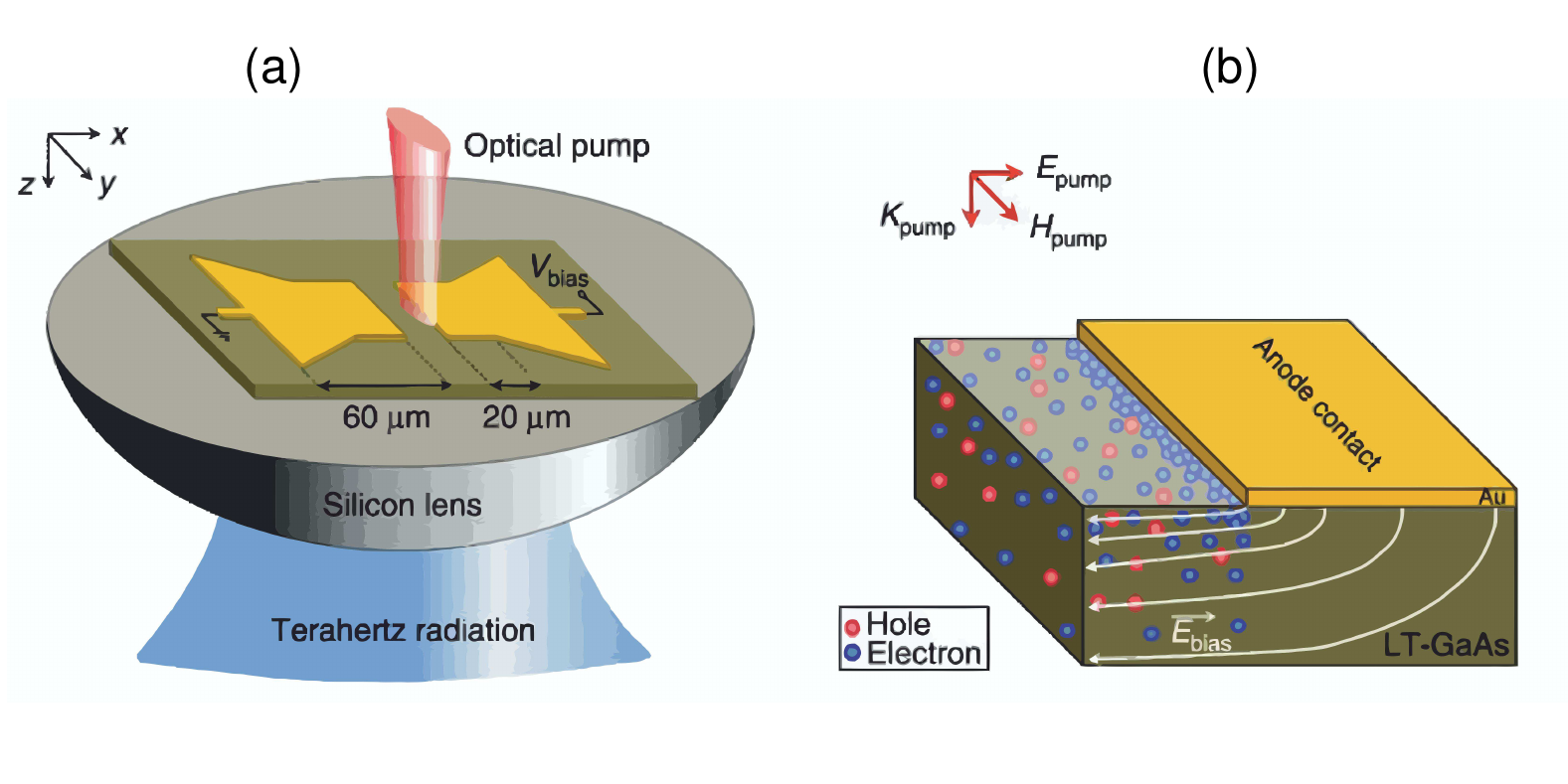}
\caption{(a) A photoconductive antenna consisting of a GaAs substrate and deposited electrodes in the bow-tie shape. (b) the distribution of the photoinduced electrons near the anode when bias voltage is applied to the electrodes.~\cite{electrodes2}}
\label{10}
\end{figure}

Authors of Ref.~\cite{resist1} have observed that various pump of the semiconductor in the THz photoconductive antenna gap can affect the output power. Moreover, there are two options for photoconductor pump: uniform illumination of the antenna gap, and partial and asymmetric illumination of the gap close to the anode (Figure~\ref{10}(a)). In the case of uniform illumination, the current of photoinduced charge carriers is directly proportional to the optical power of the pump. However, in the case of asymmetric illumination, photocurrent and hence THz signal are nonlinearly proportional to the optical pump power. This pumping regime is characterised by the increased electric field near the anode, and, consequently, increased electrons concentration in this region (Figure~\ref{10}(b)). This leads to an increase of the photocurrent, resulting in an increased THz signal power. According to Ref.~\cite{resist1}, the asymmetric illumination of the photoconductive antenna gap gives more than 4-fold enhancement compared to the uniform illumination. Moreover, it has been shown that the stronger optical field is focused in the area of the anode, the more powerful THz radiation is~\cite{geom5}.

\subsection{Continuous THz generation in photomixers} \label{first3}

The process of THz heterodyne generation in semiconductors is similar to that discussed in the preceding sections, with the only difference being the pump laser regime. Under the continuous double-wavelength pump with a frequency difference lying in the THz region, electron-hole pairs are formed in the surface layer and the conductivity of the semiconductor is modulated proportionally to the electrons and holes concentration, as the pump intensity beats with the frequency corresponding to the difference between two incoming optical pump frequencies. Such devices for photoconductive CW THz generation are usually called \textit{photomixers}~\cite{Brown1993,dipole8}. Schematic layout of CW THz generation in photomixer, and principle of its operation are shown in Figure~\ref{1a}.
\begin{figure}[!t]
\includegraphics[width=0.5\textwidth]{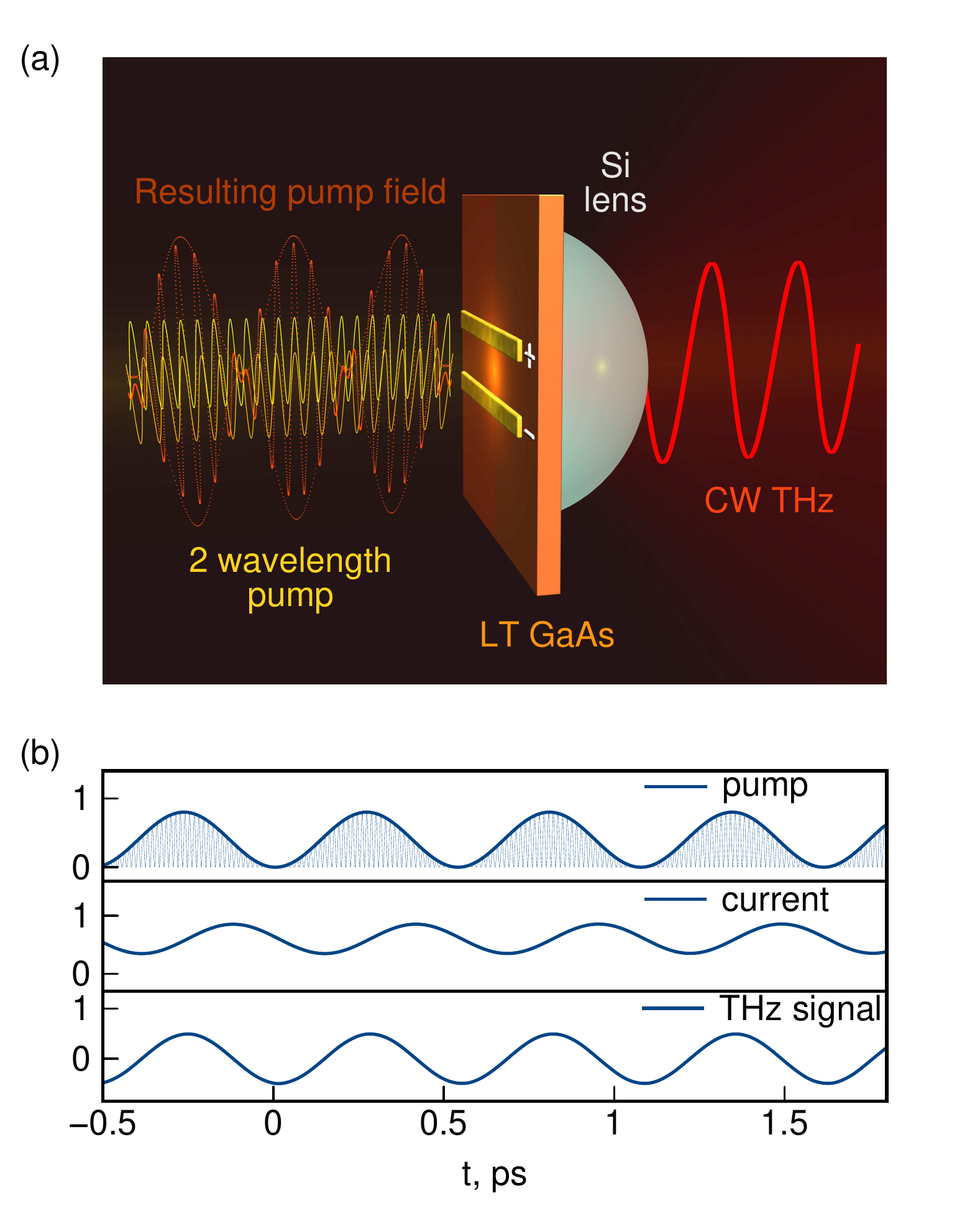}
\caption{Schematic view of a setup used to generate CW THz radiation by photoconductive LT-GaAs based antenna (a) and time evolution of pump radiation envelope, antenna photocurrent and generated THz signal (b).}
\label{1a}
\end{figure}

Surface conductivity in photomixer and charge carriers acceleration are similar to the ones in photoconductive antenna, and the number of carrier pairs similarly follows the pump evolution:
\begin{equation}
\frac{\partial n(t)}{\partial t}=\frac{\eta I_{\rm pump}\left(t\right)}{h\nu} - \frac{n}{\tau_{\rm rec}},
\label{eqpm1}
\end{equation}
where $\eta$ is the external quantum efficiency, $I_{\rm pump}\left(t\right)$ is the pump intensity, $\tau_{\rm rec}$ is the carrier recombination time, and $h\nu$ is the energy of the pump photon.

\begin{figure*}
\begin{center}
\includegraphics[width=1\textwidth]{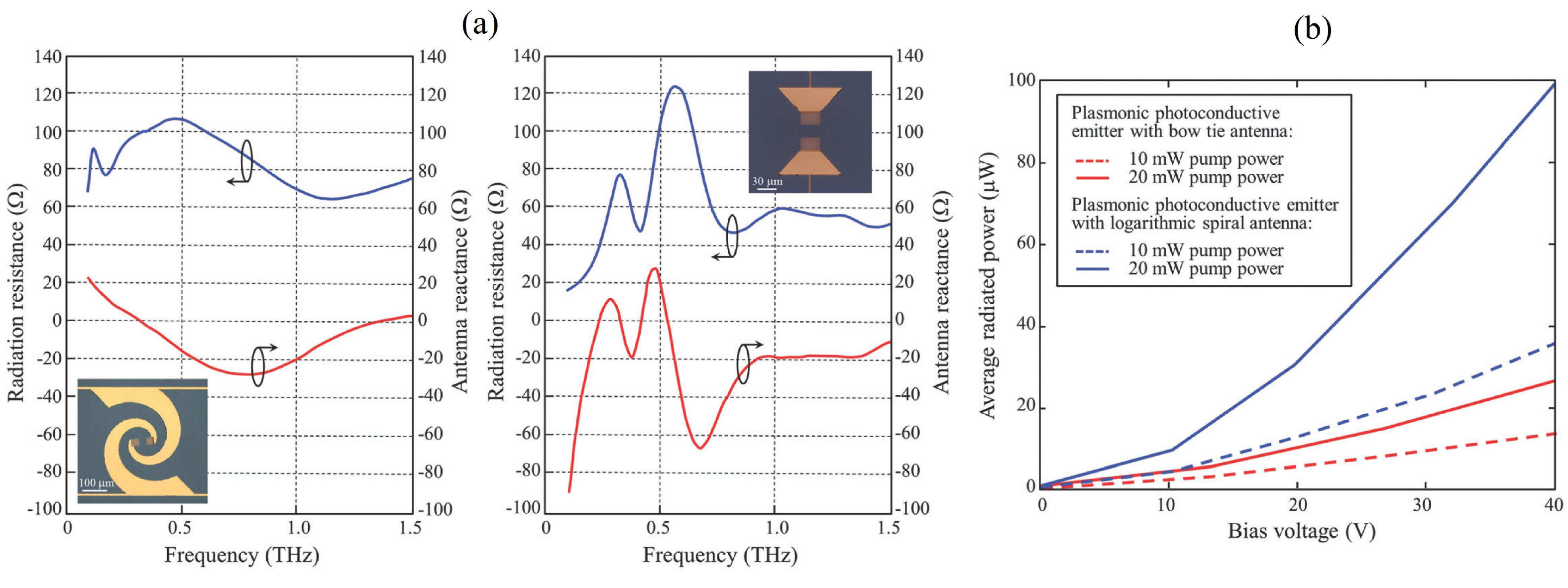}
\end{center}
\caption{(a) Plots of the frequency dependence of impedance (blue) and reactance (red) of two different THz antennas, spiral and bow-tie, correspondingly; (b) the dependencies of the radiated power of the spiral (blue) and bow-tie (red) antennas on the bias between electrodes for different powers of femtosecond pump.~\cite{resist}}
\label{omparison}
\end{figure*}

Pump radiation on the case of photomixing comprises two laser beams bringing to the surface average intensities $I_1$ and $I_2$ at the frequencies $\nu_1$ and $\nu_2$, correspondingly. Thus, the intensity at the surface would be described as:
\begin{equation}
\begin{split}
I_{\rm pump}\left(t\right)&=I_1+I_2+2\sqrt{I_1 I_2}\left(\cos{2\pi \left(\nu_1-\nu_2\right)t}\right.+\\
&+\left.\cos{2\pi \left(\nu_1+\nu_2\right)t}\right).
\end{split}
\label{eqpm2}
\end{equation}

The second oscillating term in the sum changes within a timescale that is significantly shorter than the carrier effective recombination time $\tau_{\rm c}$, hence it won't affect the photocurrent. Then after substitution of Eq.~\ref{eqpm2} into Eq.~\ref{eqpm1}, we get the evolution of carrier density~\cite{Brown1993}:
 \begin{equation}
n(t)=\frac{\eta \tau}{h\nu} \left(I_1+I_2 +  \frac{\sqrt{I_1 I_2}\sin\left(\omega t+\phi\right)}{\sqrt{1+\omega^2\tau_{\rm rec}^2}}\right),
\label{eqpm3}
\end{equation}
where $\omega=2\pi\left(\nu_1-\nu_2\right)$ is the THz frequency, and $\phi=\tan^{-1}\left(\frac{1}{\omega \tau}\right)$ in the phase delay due to the carriers lifetime.

Since the applied field is constant, for the photocurrent we obtain~\cite{Lee2009}
 \begin{equation}
J(t)=I_D+\frac{I_A}{\sqrt{1+\omega^2\tau_{\rm rec}^2}}\sin\left(\omega t+\phi\right),
\label{eqpm4}
\end{equation}
where $I_D=E_{\rm bias}\mu_e\tau_{\rm rec}\left(I_1+I_2\right)$ is the DC photocurrent and $I_A=2E_{\rm bias}\mu_e\tau_{\rm rec}\sqrt{I_1I_2}$ is the AC part of the current, responsible for the generation of the CW THz signal:
 \begin{equation}
E_{THz}\left(t\right)\propto\frac{\partial J\left(t\right)}{\partial t}=\frac{\omega I_A}{\sqrt{1+\omega^2\tau_{\rm rec}^2}}\cos\left(\omega t+\phi\right).
\label{eqpm5}
\end{equation}
It can be seen, that for the maximum conversion efficiency, the powers of both wavelength modes should be as close to each other as possible. Pump intensity profile, corresponding photocurrent and far field THz signal of the photomixer calculated with the described model are shown in Figure~\ref{1a}(b).

\subsection{Affect of the contact shape onto THz photoconductive antenna and photomixer radiative characteristics} \label{first4}

In a number of works~\cite{geom1, geom2, geom3, Jarrahi2015, frac1} it was observed that the geometry of the electrodes affects both the width of the spectrum and the position of the maximum of the radiation intensity in the far field, as well as its intensity value. Comparative analysis of different geometries was done as well. For THz antennas, the same geometries as in their RF and microwave counterparts are used: dipole, logarithmic, spiral antenna, bow-tie~\cite{resist7,resist22,resist8,resist9}.

One of the most important characteristics of all antennas, including traditional radio, microwave and THz photoconductive, is the amount of the emitted power. However, between the traditional and THz antennas there is one significant difference -- the nature of the input signal. In the case of a THz antenna it is the laser radiation that generates the nonequilibrium electrons and holes photocurrent in the surface layer of the photoconductor. Therefore, the THz photoconductive antenna, or photomixer are usually considered as a system consisting of two elements: the photoconductor and the antenna. These elements in turn are described in the traditional physical quantities, such as active and reactive conductivity, capacitance, and input impedance~\cite{resist1,optim}.
\begin{figure}[!b]
\begin{center}
\includegraphics[width=0.5\textwidth]{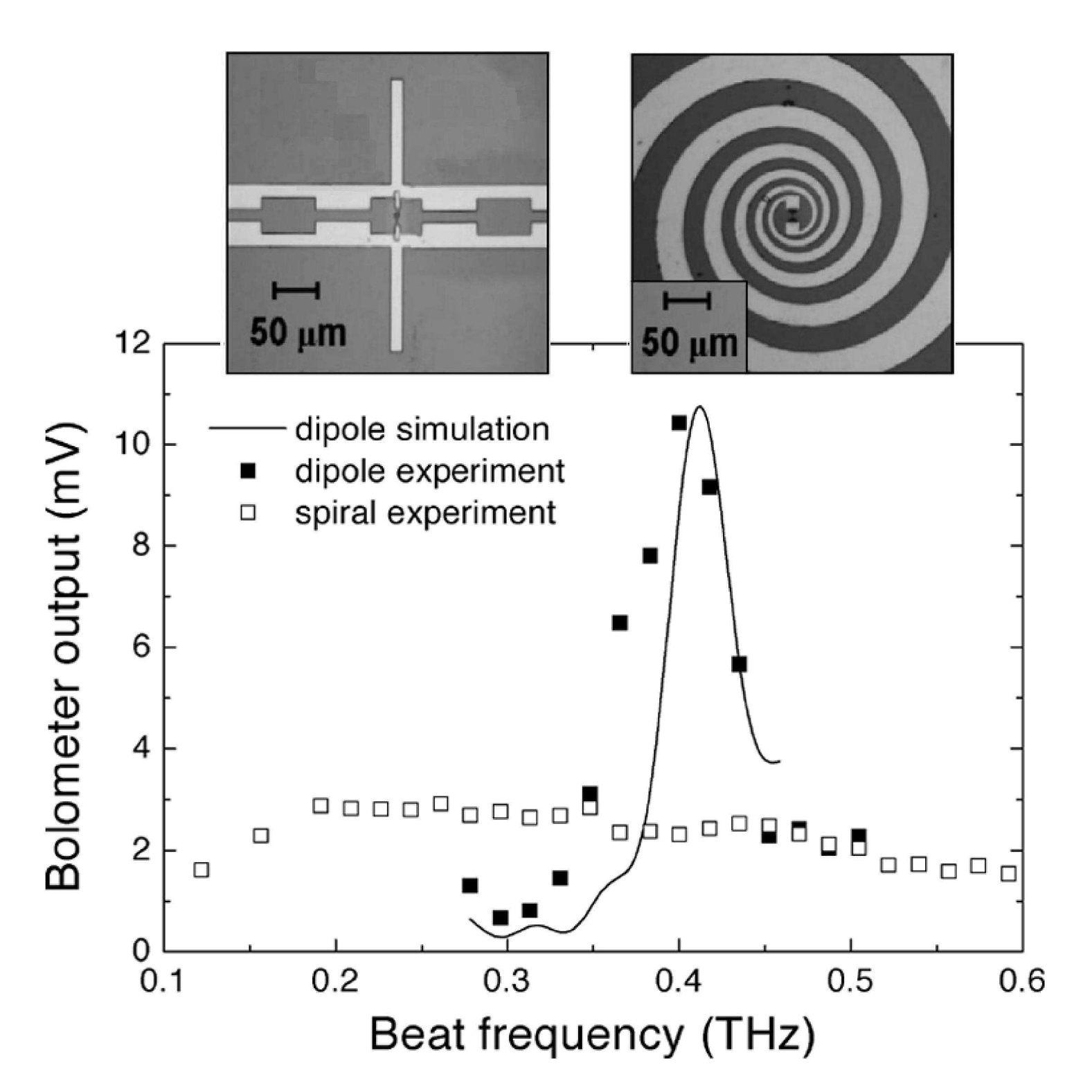}
\end{center}
\caption{Graphs of the frequency dependence of the output bolometer signal for the spiral and full-wavelength dipole THz photomixers. Dots represent experimental data, solid lines - numerical calculations.~\cite{optim}}
\label{spiraldipole}
\end{figure}

The output power $P_{\rm rad}$ of THz photoconductive antenna is given by the formula~\cite{resist1}:
\begin{equation}P_{\rm rad}(\omega) = \frac{1}{2}|S_{\rm p}(\omega)P_{\rm opt}(\omega)|^{2}\xi\frac{G_{\rm A}}{(G_{\rm A} + G_{\rm p})^{2} + (\omega C + B_{\rm A})^2},
\label{eq011}
\end{equation}
where $S_{\rm p}$ is the current susceptibility of the photoconductor, $P_{\rm opt}(\omega)$ optical pump power at frequency $\omega$, $\xi$ is the antenna efficiency (describes the dissipative losses in the antenna material~\cite{Balanis2005}), $G_{\rm A}$ and $B_{\rm A}$ are active and reactive conductances of the antenna respectively, $G_ {\rm p}$ is the conductivity of the photoconductor, and $C$ its capacitance. Regardless of the THz photoconductive antenna geometry, in order to maximise its radiated power, its stray conductance must be much less than its active conductance$(\omega C + B_{\rm A}) \ll (G_{\rm A} + G_{\rm p})$. Full or partial compensation of reactive conductance can be achieved by maintaining inductive load of the antenna on the same level as capacitive load of the photoconductor, or by adding a negative reactive conductance shunt into the circuit~\cite{dipole9, resist13}.

In Ref.~\cite{resist} a comparative analysis of two common THz antenna geometries is performed: spiral and bow-tie. Figure~\ref{omparison} shows plots of the dependence of radiation resistance and reactance of these antennas on the frequency and the dependence of the radiated power on the bias applied to electrodes for THz frequency corresponding to the maximum output power (Figure~\ref{omparison}(b)). It can be seen that the spiral antenna having larger radiation resistance and smaller reactance, emits more effectively than the bow-tie within wider range of frequencies. Thus, the maximum active resistance and the minimum absolute value of the reactive resistance are prerequisite for efficient THz radiation generation~\cite{resist22, resist24}.

Since active and reactive resistances affect the power emitted by THz antenna, the spectral characteristics of antenna impedance and reactance determine the spectral characteristics of the THz signal. It has been observed that antenna with a radiation resistance, weakly dependent on frequency in the THz range, has a wider frequency range of effective operation~\cite{resist6, resist}. Spiral and a bow-tie geometries are examples of such antennas. Figure~\ref{spiraldipole} shows the spectrum of THz signal from the spiral photomixer antenna measured by bolometer. At frequencies between 0.1 and 0.6~THz the signal power is almost constant~\cite{optim}. On the other hand, antennas with a pronounced resonance characteristic impedance of, for example, full-wavelength dipole antenna have narrow-band emission spectrum (Figure~\ref{spiraldipole})~\cite{dipole9, resist4, resist5}.

In the work~\cite{frac1} bow-tie antenna are studied, made in the form of Sierpinski triangles, the results are shown in Figure~\ref{Fract}. After analyzing them, one can conclude that increasing the order of fractality allows gaining the power emitted by the bow-tie to the far field.
\begin{figure}
\begin{center}
\includegraphics[width=0.5\textwidth]{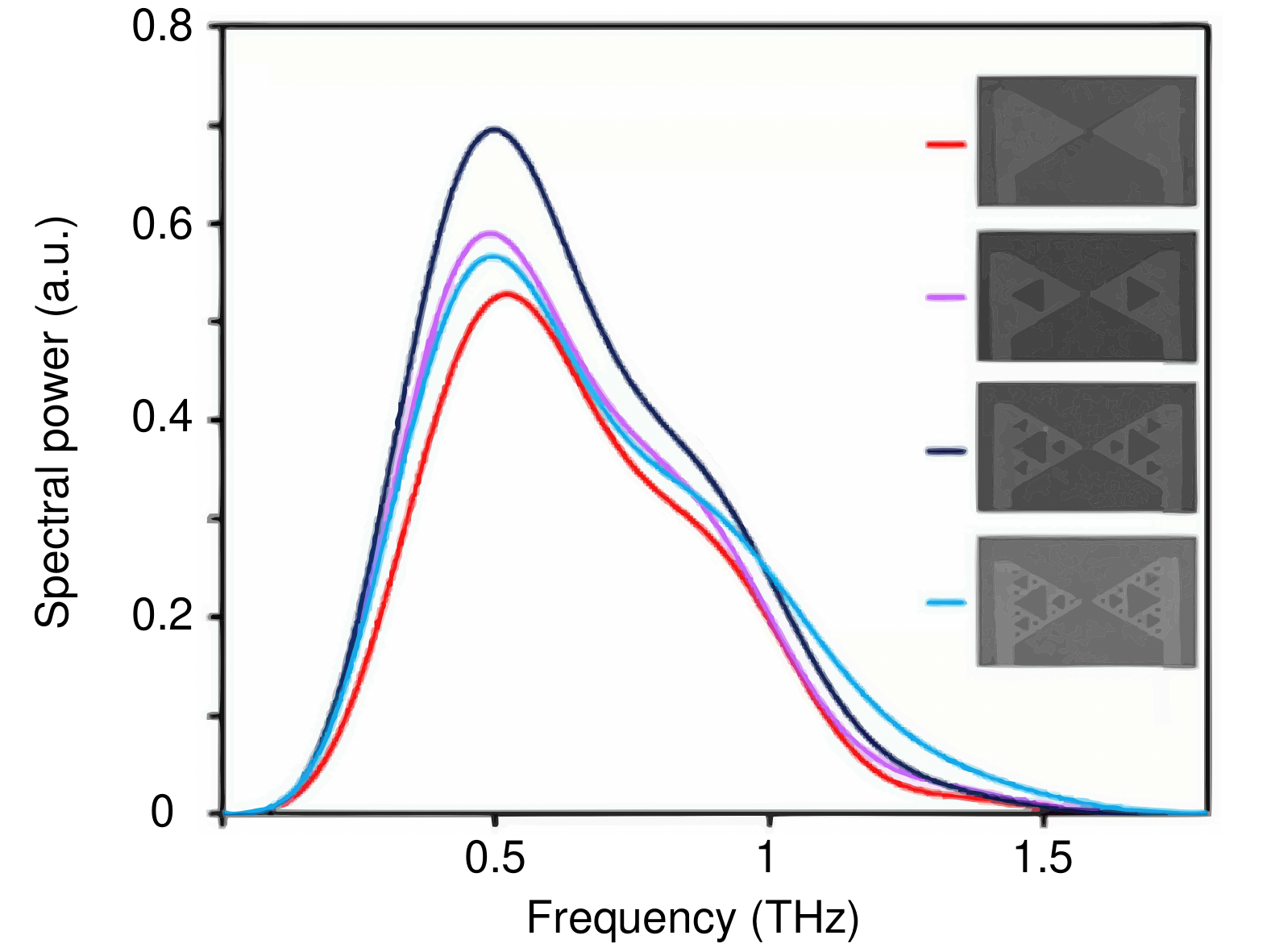}
\end{center}
\caption{Curves of THz power spectra for different bow-tie fractal antennas.~\cite{frac1}}
\label{Fract}
\end{figure}

\section{Hybrid Photoconductive Antennae for Terahertz Generation} \label{second}

\subsection{{Principle of a hybrid THz-optical photoconductive antennas operation}} \label{second01}

Availability of powerful optical sources of ultrashort light pulses, as well as technologies of manufacturing the necessary semiconductor materials and deposition of metal structures on them makes photoconductive antennas promising sources of THz waves. However, the effectiveness of traditional photoconductive antennas is limited by low drift velocities of the photoinduced charge carriers in semiconductor substrates~\cite{3d42, 3d43, carrier1}. For efficient optical-to-terahertz conversion in the frequency range above 0.5~THz a relaxation time of carriers $\tau_{\rm rel}$, as discussed earlier in Section~\ref{first} must be subpicosecond. According to the formula:
\begin{equation}l = v_{\rm e,p}\tau_{\rm rel}\label{eq12},\end{equation}
the mean free path \begin{math}l\end{math} is about 100~nm. Therefore, only a small part (about 5\%~\cite{electrodes2}) of nonequilibrium electrons and holes reaches the electrodes and participate in the efficient THz generation. The rest of the charge carriers recombines with charges of opposite sign on the way towards the electrodes. This effect is called \textit{carrier screening effect}~\cite{3d42, carrier1}. The diffraction limit for electromagnetic waves does not allow to focus the laser pump into the spot of 100~nm, to allow involvement of all the excited carriers. Therefore, in order to reduce the effect of screening and to improve the properties of photoconductive antennas, optical nanoantennas are used~\cite{optic1}.

Optical nanoantennas possess all the inherent characteristics of their radio and THz counterparts, except of course that they operate on electromagnetic waves with frequencies in the optical spectral range, whereby the dimensions of such antennas are typically of the order of several hundreds of nanometers. Optical nanoantennas are used extensively in devices that use photon-electron conversion, such as photodiodes, LEDs and photovoltaic converters~\cite{optic2, optic3}. Nanoantennas can be plasmonic (metal), dielectric and metal-dielectric~\cite{optic2, optic3}. The most widely used and well-studied nanoantennas are the plasmonic ones. Recently, a number of papers appeared that demonstrate the successful enhancement of optical-to-terahertz conversion as a result of the integration of such plasmonic antennas into the gap between metal electrodes of THz photoconductive antennas to enhance the absorption cross-sections of laser radiation into semiconductor substrate. Such a device (which combines optical nanoantennas and THz photoconductive antenna) is known as a hybrid optical-to-THz photoconductive antenna.

Receiver optical nanoantennas, as part of a hybrid THz-optical photoconductive antenna or photomixer, convert the pump optical radiation into a strong near field and redistribute energy into the semiconductor surface layer. Local electric field enhancement is described by the local field enhancement factor:
{\begin{equation}\delta = \frac{|\boldsymbol{E}|}{|\boldsymbol{E}_{\rm 0}|}.
\label{eq122}
\end{equation}
This value shows what fold increase gets the module of the electric field at the point $|\boldsymbol{E}|$ in the presence of the nanoantenna compared to the one without nanoantenna $|\boldsymbol{E}_{\rm 0}|$.}

{Local electric field enhancement occurs due to the structure plasmon resonance~\cite{plasm3,geom3,electrodes1}. At the same time, surface plasmon excited in the metal structure asymmetrically scatters the incident wave, and the field is redistributed in the semiconductor substrate so that the local electric field enhancement factor near the electrodes becomes much greater than 1. Therefore, the optical power absorbed by the semiconductor $P_{\rm opt}$ in the volume $V$:}
\begin{equation}P_{\rm opt} = \frac{1}{2}\int\sigma{\delta}^2|\boldsymbol{E}_{\rm 0}|^2dV \label{eq123}\end{equation} increases proportionally $\delta^2$. Because of this, the concentration of nonequilibrium charge carriers near the electrodes $n_{\rm e,p}$ increases, and, according to the dependence:
\begin{equation}E_{_{\rm THz}} \propto \frac{d(q_{\rm e,p}v_{\rm e,p}n_{\rm e,p})}{dt}, \label{eq13}\end{equation}
where $E_{_{\rm THz}}$ is the amplitude of the output THz signal, $t$ is the process duration, the efficiency of optical-to-terahertz conversion increases. It is important to note that physical dimensions of the nanoantennas and the material they are made of play an important role in the excitation of plasmon resonance.

Moreover, optical nanoantennas embedded in a THz photoconductive antenna, improve thermal stability of the latter, due to the large thermal conductivity of gold, which they are made of~\cite{gex1}. The excess of heat in traditional THz antennas leads to the thermal generation of carriers after optical pulse excitation. Recombination of thermal and photoinduced carriers in subsequent cycles of charge carriers generation prevents the stable operation of the photoconductive antenna. Thus, higher thermal efficiency also makes optical nanoantennas attractive for use in THz sources.

Thus, the design of optical nanoantennas embedded in a THz photoconductive antenna should maximize the transfer of light energy into the absorptive semiconductor substrate and minimize the path length of charge carriers traveling between metallic electrodes. Here, we review the most recent works and organise them according to the approach to optical nanoantenna realisation.

\subsection{Plasmon Monopoles} \label{second3}

The simplest approach to realisation of optical nanoantennas is monopole nanoantennas. Monopole nanoantennas are able to provide strong localisation of laser radiation power near them~\cite{optic2}. The characteristics of such antennas are highly dependent on the size and material of the nanoparticle, but still have the least number of tunable parameters and are therefore particularly important for applications that require the large nanoantenna array, for example in the gap of the THz photoconductive antenna.
\begin{figure}
\begin{center}
\includegraphics[width=0.5\textwidth]{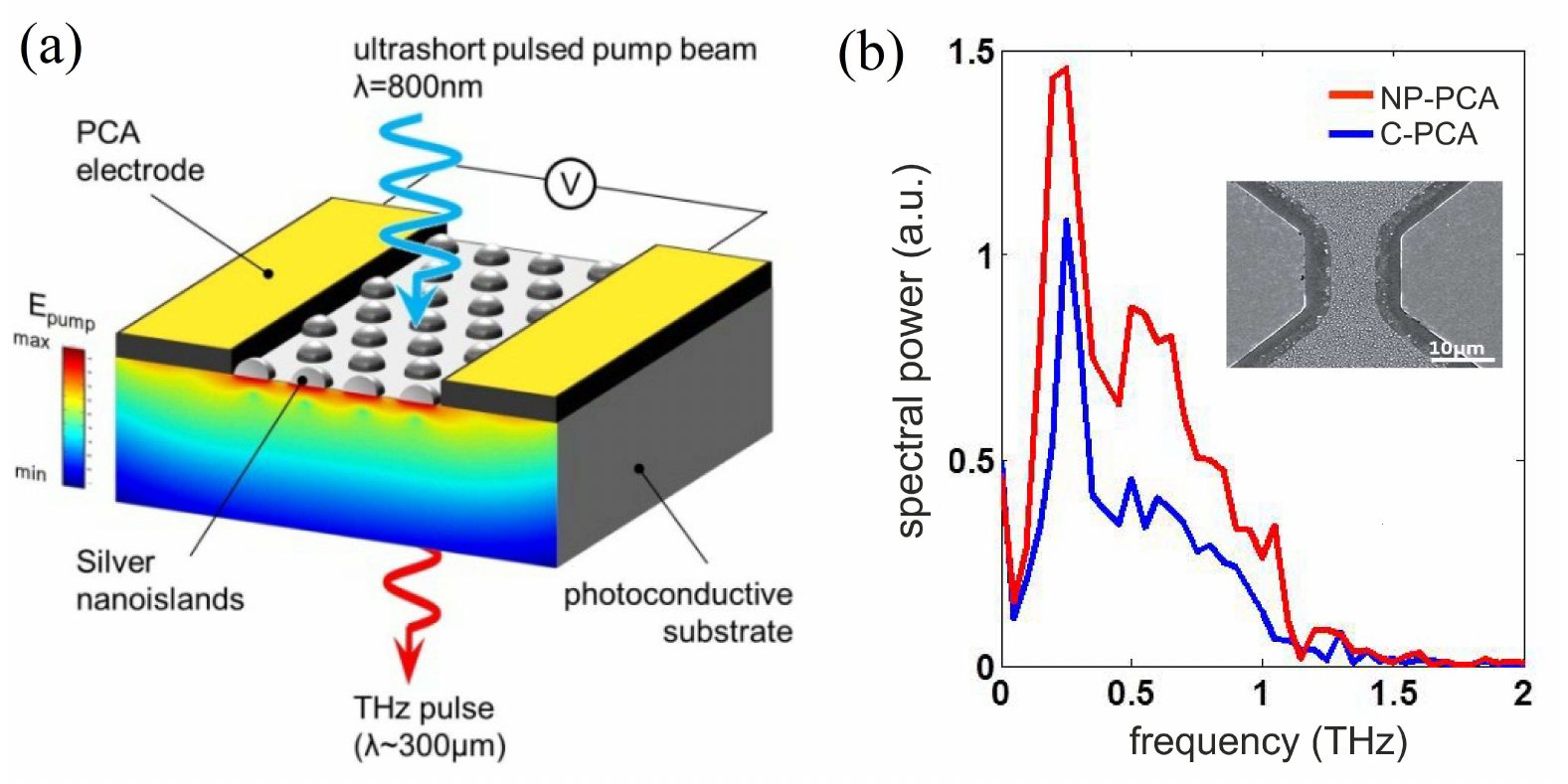}
\end{center}
\caption{(a) Schematic representation of the dipole photoconductive THz antenna with a large area of excitation in the gap with metal nanoislands. (b) Plots of THz spectral power for the traditional (C-PCA) and hybrid (NP-PCA) THz photoconductive antenna and SEM image of silver nanoislands.~\cite{mono1}}
\label{mono}
\end{figure}

The authors of~\cite{mono1} propose silver plasmonic monopole nanoantennas in the form of so-called \textit{nanoislands} (Figure~\ref{mono}(a)). The fabrication of such monopole nanoantennas is done by thermal evaporation of silver (Ag), with subsequent deposition onto LT-GaAs antenna substrate surface. This method can be used to pattern nanostructures inside a large area, because of lower costs if compared to, for example, electron beam lithography used for the precise fabrication of metal nanostructures~\cite{mono2}. Silver in the hybrid THz-optical antenna is used due to the high quality factor of the silver nanostructures in the visible range, which allows achieving maximum localization of the field in the semiconductor surface layer in close proximity to a nanoantenna, with minimal losses. In addition, Ag has a relatively low melting temperature, and therefore its thermal evaporation is possible without thermal damage to the semiconductor GaAs substrate~\cite{mono4}. The geometrical parameters of the monopoles are specially derived so that surface plasmon waves are excited in the structure under an optical pump. For a pump radiation of 800~nm wavelength, which is typical for Ti:Sapphire femtosecond lasers, nanoislands should have a thickness of 20~nm and a diameter of 173~nm.
\begin{figure}[!t]
\begin{center}
\includegraphics[width=0.5\textwidth]{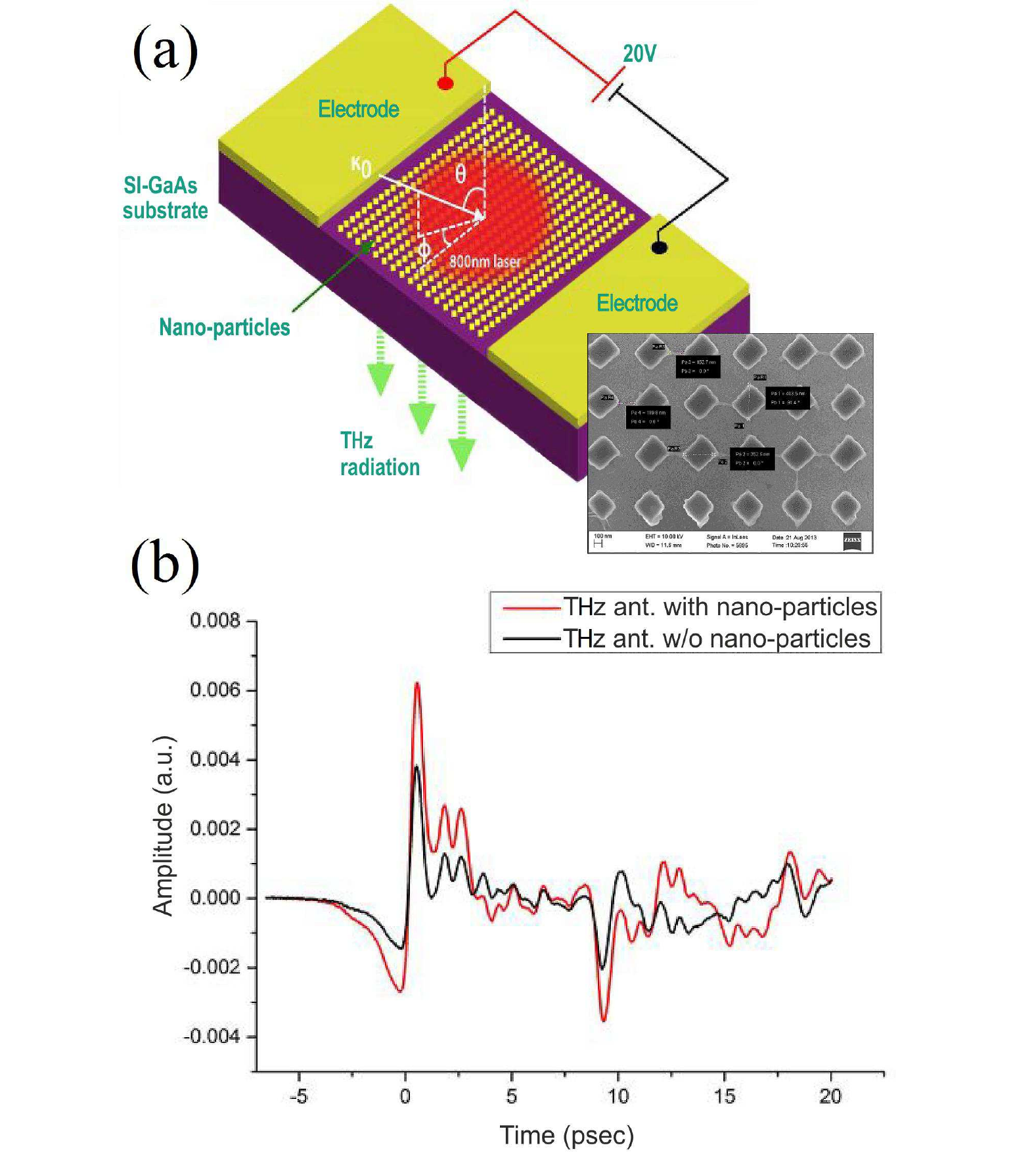}
\end{center}
\caption{(a) Hybrid THz-optical photoconductive antenna with an array of monopole nanoantennas, the array is fabricated on a substrate of semi-insulating GaAs (SI-GaAs) (b) Output signal of THz photoconductive antennas with(red) and without (black) monopole nanostructures.~\cite{monop}}
\label{monop}
\end{figure}
\begin{figure*}[]
\begin{center}
\includegraphics[width=1\textwidth]{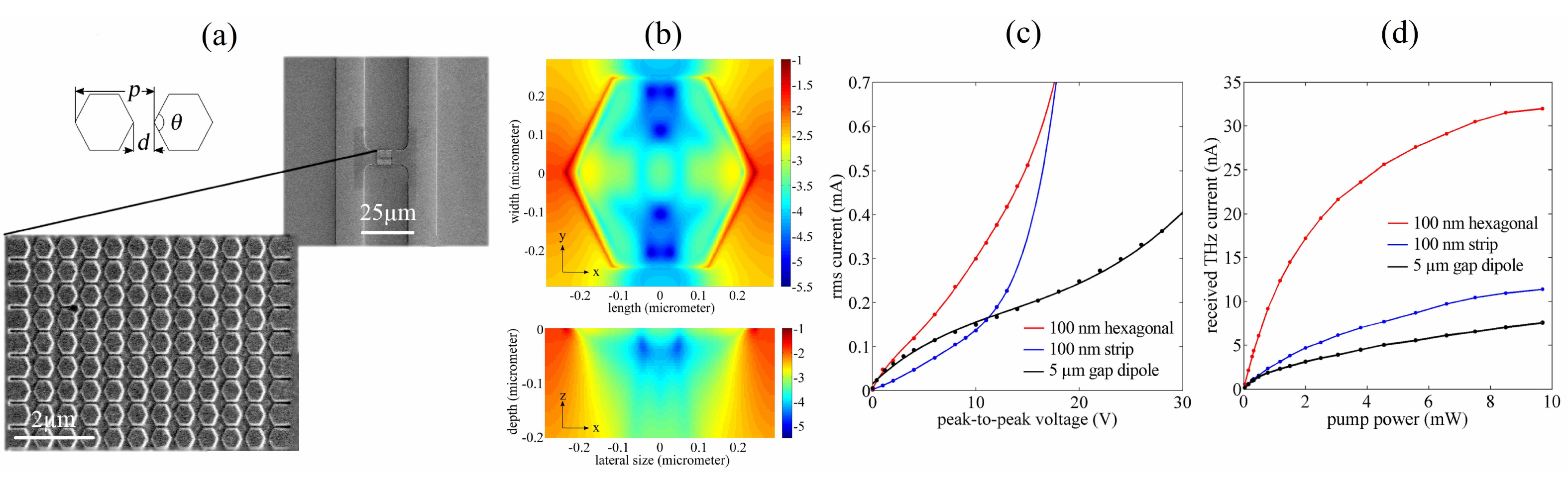}
\end{center}
\caption{(a) Image of a hexagonal array nanoantenna on SI-GaAs substrate, obtained using scanning electron microscopy. (b) Picture of optical field distribution near the hexagonal nanoantenna. (c) Peak-to-peak voltage dependance of photo-excited current in the hybrid THz antennas with hexagonal monopoles (red curve), strip plasmonic gratings (blue curve) and conventional dipole antenna. (d) Comparison of received THz currents from the hybrid THz antennas with hexagonal monopoles, strip plasmonic gratings and conventional dipole antenna depending on pump optical power.~\cite{gex1}}
\label{gexa1}
\end{figure*}

Figure~\ref{mono}(b) shows the comparative characteristics of traditional (C-PCA) and nanoisland improved (NP-PCA) THz photoconductive antenna. It is clearly seen that the hybrid structure has 1.5-2 times higher conversion efficiency, and particularly enhances high-frequency part of the generated spectrum. This advantage is explained by the following dependence (Eq.~\ref{eq13}): monopole nanoantenna concentrate the light inside the gaps between adjacent nanoparticles, so that the absorption of optical radiation in these areas increases, hence grows the concentration of nonequilibrium carriers $n_{e,p}$, which, ultimately, increases the amplitude of the output THz signal $E_{_{\rm THz}}$. However, this field redistribution leads to a decrease in its value under nanoislands and significantly reduces the density of electrons and holes in these areas. However, this does not affect the total current of photoinduced charges, as silver monopoles are almost perfect conductors of an external electric field. This process is described in detail in the following works:~\cite{mono1,optic1}.

Another possible realization is the array of rectangular monopole nanoantennas inside the photoconductive gap of the photoconductive antennas and photomixers~\cite{monop}. Figure~\ref{monop}(a) schematically shows a THz antenna with array of rectangular nanostructures and SEM image of these nanostructures. The mechanism of optical field local enhancement in these nanoantennas is similar to the one in previously described - plasmon resonance excitation. Metal deposition followed by electron beam lithography is used for the fabrication of such monopole nanoantennas. However, instead of Au metal nanostructures, which fabrication is a very expensive process, arrays of gold and germanium (AuGe) alloy monopoles are used, that not only greatly reduce costs but also reduce the risk of the thermal damage to SI-GaAs at thermal evaporation of the metal due to the relatively low melting temperature of AuGe. These nanostructured hybrid THz-optical photoconductive antennas are able to double the generation of THz radiation in comparison with traditional photoconductive antennas (Figure~\ref{monop}(b)).

Paper~\cite{gex1} describes hexagonal metal nanostructures superior to plasmonic nanorods (see below) in THz generation enhancement. Figure~\ref{gexa1}(a) shows the array of these monopole nanoantenna inside the photoconductive gap of the dipole THz PCA. Such structures are fabricated with the focused ion beam technology. It has been shown, that an important property of THz photoconductive antennas with integrated hexagonal nanoantennas is their \textit{high resistance to thermal breakdown}. The increase of pumping intensity leads to an increase of photoinduced current carriers, which leads to heating of the semiconductor substrate and its thermal breakdown~\cite{gex40,gex41}. Figure~\ref{gexa1}(c) demonstrates the advantage of the hexagonal nanoantennas over grating nanoantennas and a traditional THz antennas. It is seen that to achieve similar photoinduced carriers current values, different antennas require application of different bias between the electrodes. Moreover, maintaining a predetermined value of the current in the structure with a hexagonal monopoles requires less external stress than in other antenna realisations. This suggests that traditional antennas and plasmonic gratings generate more heat than the hexagonal nanoantenna, and, accordingly, have a greater risk of thermal breakdown~\cite{gex1}.

Similar comparison is shown in Figure~\ref{gexa1}(d), but for the THz power dependence on the average power of femtosecond pump laser. When comparing grating and hexagonal nanoantenna, the obvious superiority of the latter is clearly seen. This is due to a combination of the external field generated by the electrodes, and localisation of photoinduced charges in the proximity to the vertices of hexagons (Figure~\ref{gexa1}(b)).
\begin{figure}[!b]
\begin{center}
\includegraphics[width=0.5\textwidth]{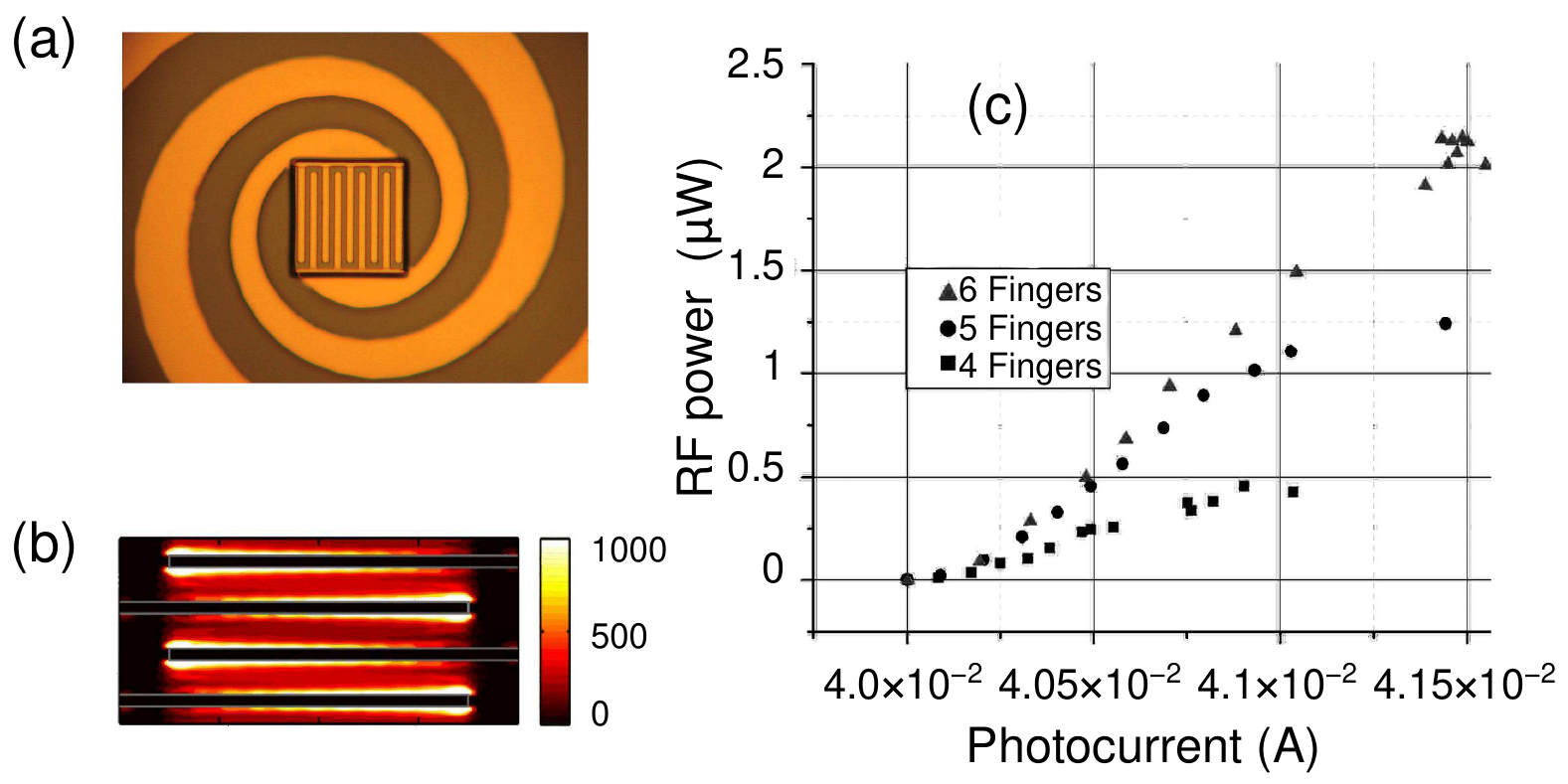}
\end{center}
\caption{(a) Image of the spiral THz photomixer with built-in finger nanoantenna. (b) Electric field distribution in finger nanostructure of THz photomixer. (C) Comparison of the radiated power in nanostructured THz photomixers with different number of fingers.~\cite{dipole10,dipole11,dipole7}}
\label{dipolefin}
\end{figure}
\begin{figure*}
\begin{center}
\includegraphics[width=0.95\textwidth]{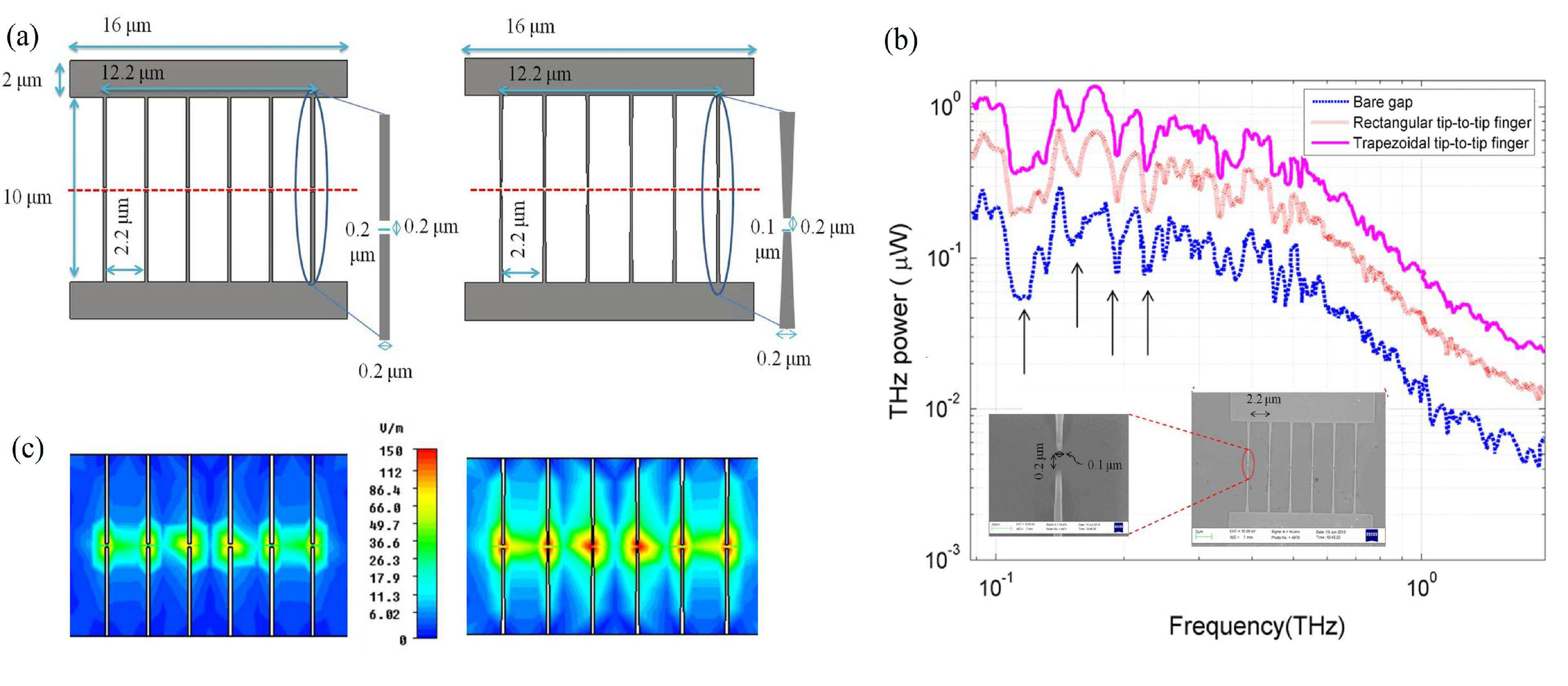}
\end{center}
\caption{Structural scheme (a) and the distribution pattern of the optical field (b) of rectangular (left) and trapezoidal (right) dipole nanoantennas embedded in a THz photoconductive antenna. (c) The plot of THz radiation power and frequency for hybrid-optical photoconductive dipole antennas with built-in trapezoidal and rectangular nanoantennas and traditional THz photoconductive antenna.~\cite{dipole1}}
\label{dipole}
\end{figure*}

\subsection{Metal dipole nanoantennas} \label{second0}

Typical representatives of both metal and dielectric nanoantennas are dipole nanoantennas~\cite{optic4}. As the name implies, they represent a dipole structure consisting of two elongated nanoparticles. Such structures, especially metal, when exposed to an external electromagnetic field provide strong field localization in the gap between the nanoparticles, and a large value of the absorption cross section of quantum detector placed in the gap. Due to this property, the metal dipole nanoantennas are used in THz photoconductive antennas and photomixers, enhancing an optical field of pump laser and redistributing it into the surface layer of the semiconductor substrate near the electrodes.

The first implementation of dipole nanoantennas embedded inside the gap of THz antenna are interdigitated nanostructured electrodes, or as they often called, "nano-fingers"~\cite{dipole4,dipole5,geom2,dipole2,dipole8,dipole9}. Figure~\ref{dipolefin}(a) presents a interdigitated nanostructures embedded in the gap between the electrodes of the logarithmic photomixer. THz photomixers with such nanostructures feature lower parasitic capacitance and, consequently, lower reactance if compared with traditional THz photomixers, as well as higher optical field localization inside the working area of the antenna (see Figure~\ref{dipolefin}(b)). Hence, photomixers with nanostructured dipole antennas in the photoconductive gap are more effective sources of CW THz radiation~\cite{optim} if compared to microstructured ones. In some works~\cite{dipole10,dipole11}, it has been shown that with increasing number of dipoles (fingers), the optical-to-THz conversion efficiency increases (Figure~\ref{dipolefin}(c)). This dependence can be explained by the smaller distance between the dipoles in THz photomixer with greater number of fingers, hence increased degree of the optical field localization in the gap and decreased effective lifetime of photoinduced charge carriers.

An updated approach to the design of the embedded dipole nanoantennas are the so-called dipole tip-to-tip nanoantennas~\cite{dipole1,dipole7,dipole2}. Two most commonly used realizations of such nanostructures embedded in a THz photomixer are shown above. On one side traditional rectangular dipole nanoparticles are shown and trapezoidal bow-tie nanoantennas on the other (see Figure~\ref{dipole}(a)). Being a continuation of the metal electrodes, these tip-to-tip nanoantennas are designed to reduce the path length of photoinduced electrons and holes~\cite{optic4}. Moreover, due to the nanoscale gap between the dipoles, tip-to-tip nanoantennas enhance the localization of the external optical field~\cite{dipole3,dipole2,dipole7}. High localization of the field in the gap (see Figure~\ref{dipole}(c), left picture) occurs because of the surface plasmon excitation due to plasmon resonance at the antenna excitation wavelength~\cite{optic4}. This increases a laser radiation absorption in the semiconductor substrate within the gap, and, consequently, the concentration of charge carriers in the gap. The rapid increase in the charge carriers concentration leads to a huge impulse of the photocurrent that is much greater than in conventional THz antennas. The amplitude of THz radiation is directly proportional to the rate of change of the photocurrent density, according to Eq.~\ref{eq13}, so THz dipole antenna with integrated optical nanoantennas demonstrate higher output power of THz signal than traditional one at the same pump rate, only because of such nanoscale extensions of the electrodes (Figure~\ref{dipole}(b)).

In Ref.~\cite{dipole1} it has been noticed, that due to lower capacitive reactive resistance, receiving ability of the bow-tie antenna is much better than one of the rectangular-shaped nanoantennas, as a result, the degree of the optical field localization in the trapezoidal nanoantenna gap is significantly higher (Figure~\ref{dipole} (c), right picture)~\cite{dipole1}. In addition, a smaller capacitance increases the efficiency of generation of THz radiation by such antenna.~\cite{optim}.

\subsection{Two-dimensional plasmonic gratings} \label{second1}

Numerous researchers~\cite{optic1, electrodes2,resist,Jarrahi2015,geom3,grate1,geom2, mona} have shown that on of the most effective implementation of dipole nanoantennas in the THz photoconductive antennas and photomixers are arrays of metal dipoles whose length is of the order of the gap size between the electrodes. Arrays of dipole plasmonic nanoantennas have received the name of plasmonic gratings or plasmonic nanorod gratings.~\cite{optic1}.
\begin{figure} [b]
\begin{center}
\includegraphics[width=0.5\textwidth]{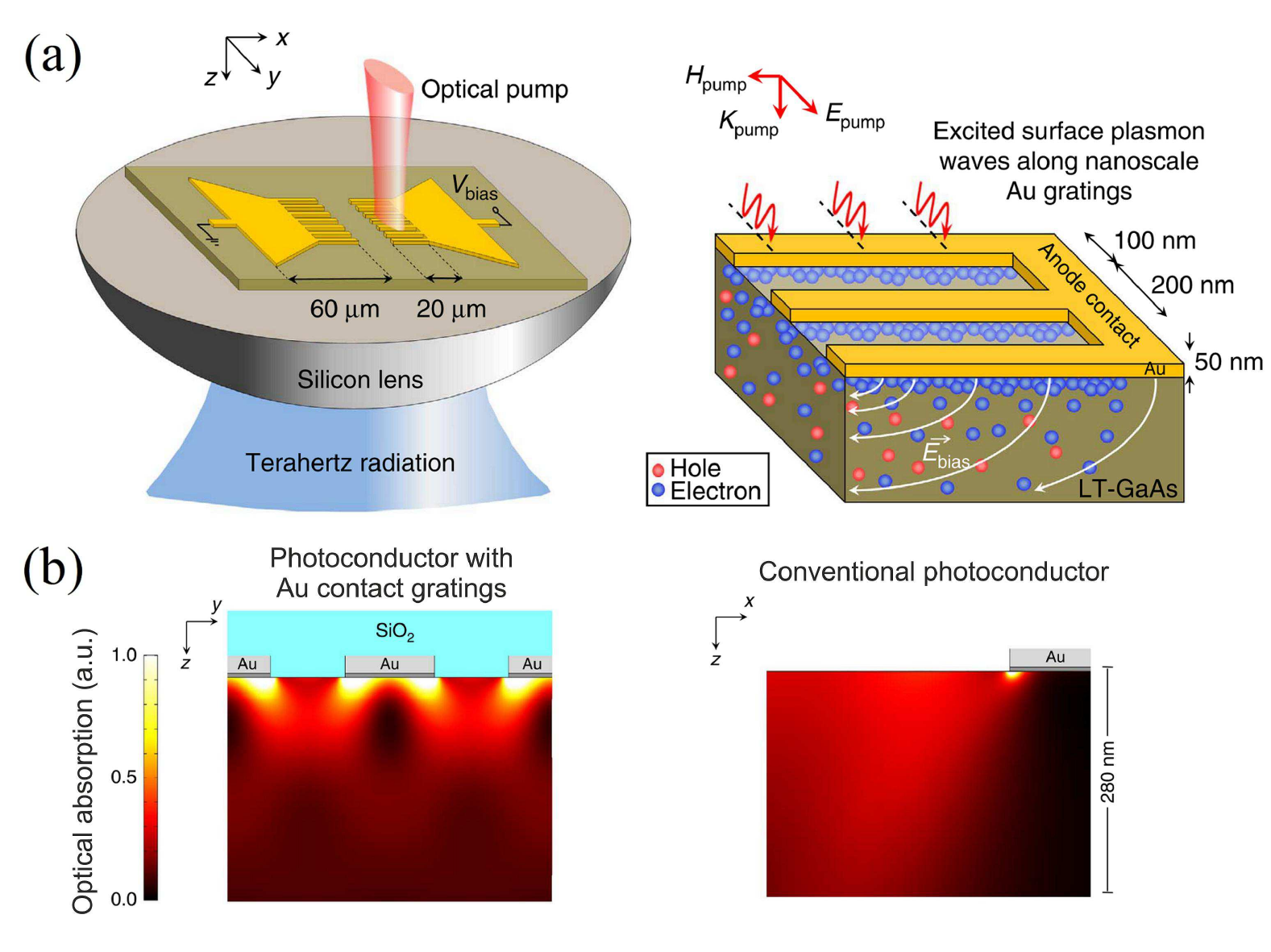}
\end{center}
\caption{(a) The geometry of the hybrid THz-optical photoconductive antenna and the distribution of electrons and holes near plasmonic nanorods. (b) The image of the distribution of absorbed optical energy in GaAs substrates with plasmonic gratings on the surface and standard THz photoconductive antenna~\cite{electrodes2}. }
\label{101}
 \end{figure}

Figure~\ref{101}(a) shows the bow-tie THz photoconductive antenna with embedded plasmonic gratings. Usually the penetration of light through the apertures is limited by the diffraction limit. However, the geometrical parameters of the gratings can be chosen so that the radiation of a femtosecond laser with a wavelength of 800~nm excites the surface plasmon wave in this periodic structure. The excitation of plasmon waves allows the transfer of the most part of optical pump energy into a semiconductor substrate~\cite{plasm2}. The surface plasmon greatly enhance the intensity of the optical field near the metallic electrodes (Figure~\ref{101}(b)), as a result, the path length of photoinduced electrons and holes to the anode and cathode is reduced in comparison with traditional THz antennas (Figure~\ref{1}).

Plasmon resonance in this structure is very sensitive to the polarization of the incident light. When the electric field of the optical pulse is parallel to the plasmonic lattice nanorods, the reflection from the grating surface is sufficiently large. On the contrary, in the case of perpendicular orientation of the electric field vector and nanorods, maximum conversion of incident light into plasmon waves is achieved~\cite{optic1}.

Figure~\ref{grate} depicts another possible implementation of plasmonic nanorod gratings embedded in dipole photoconductive THz antenna. Instead of comb-like plasmonic grating structures described above, the ends of plasmonic nanorods in this configuration are connected together, which significantly shortens the path of the charge carriers to the electrodes. Moreover, this design allows to reduce the capacitive reactance of the THz antenna that improves its emitting ability and increases antenna gain coefficient. A disadvantage of this plasmonic grating realization is that only half of nonequilibrium electrons and holes reach the electrodes within subpicosecond time. This limits the effectiveness of such structures to 50 percent~\cite{electrodes1}.
\begin{figure}[!t]
\begin{center}
\includegraphics[width=0.5\textwidth]{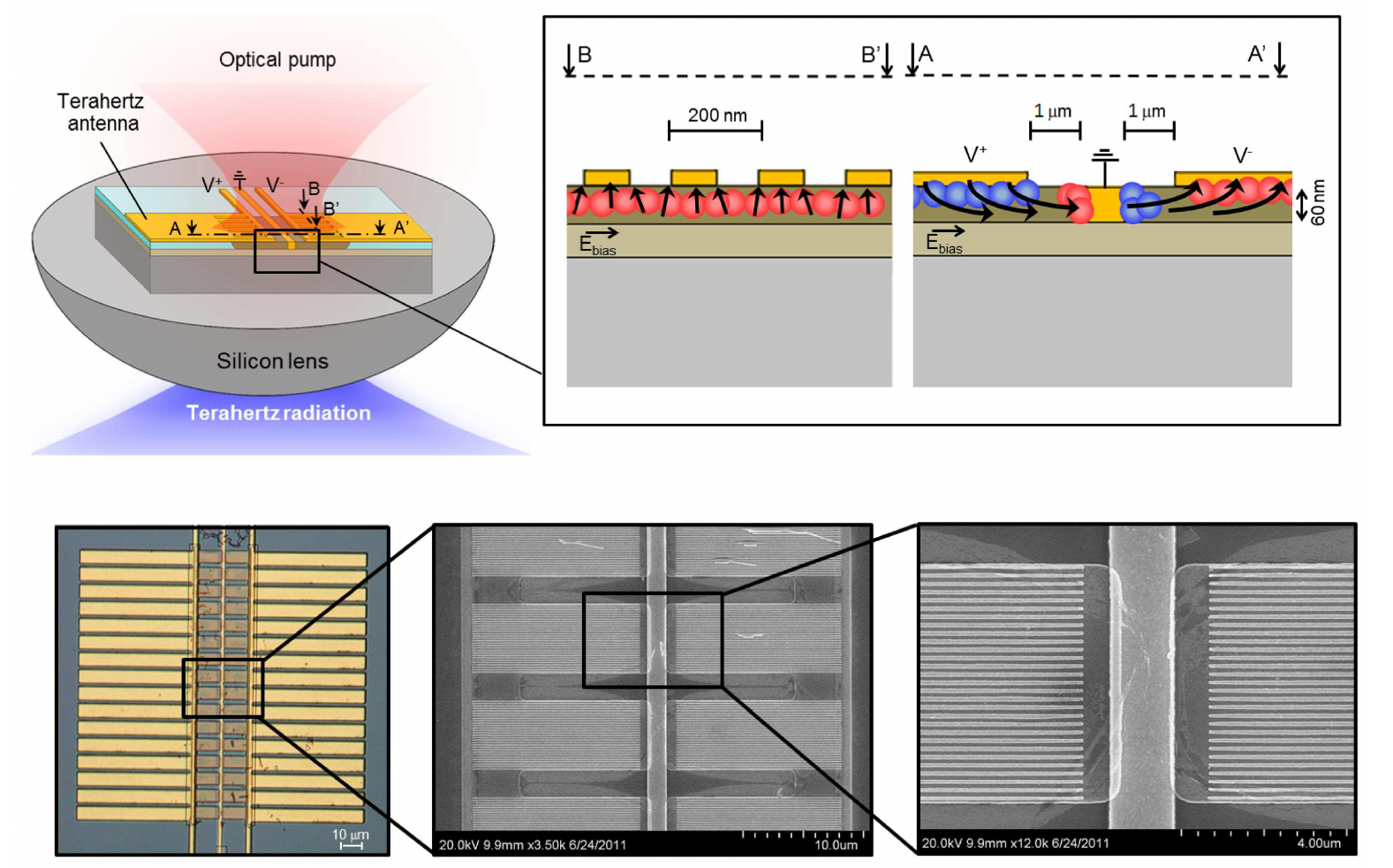}
\end{center}
\caption{Structural diagram of the THz photoconductive antenna with embedded plasmonic nanorods and ground electrode, its principle of operation (top). Top view of an array of such antennas and their SEM image (bottom).~\cite{electrodes1}}
\label{grate}
\end{figure}

In order to maintain the operation of the hybrid THz-optical photoconductive antenna without reducing its effectiveness, midway between the anode and the cathode grounding electrode is placed. This electrode is usually embedded into the semiconductor substrate and during a cycle of photoinduced charge carriers generation collects those electrons and holes that reach the electrodes, preventing their slow accumulation in the substrate. In the absence of a grounding electrode these unwanted carriers recombine with charges of opposite sign of the subsequent cycles of generation, thereby reducing the total number of particles that reach electrodes~\cite{electrodes1}.
\begin{figure}[!b]
\begin{center}
\includegraphics[width=0.5\textwidth]{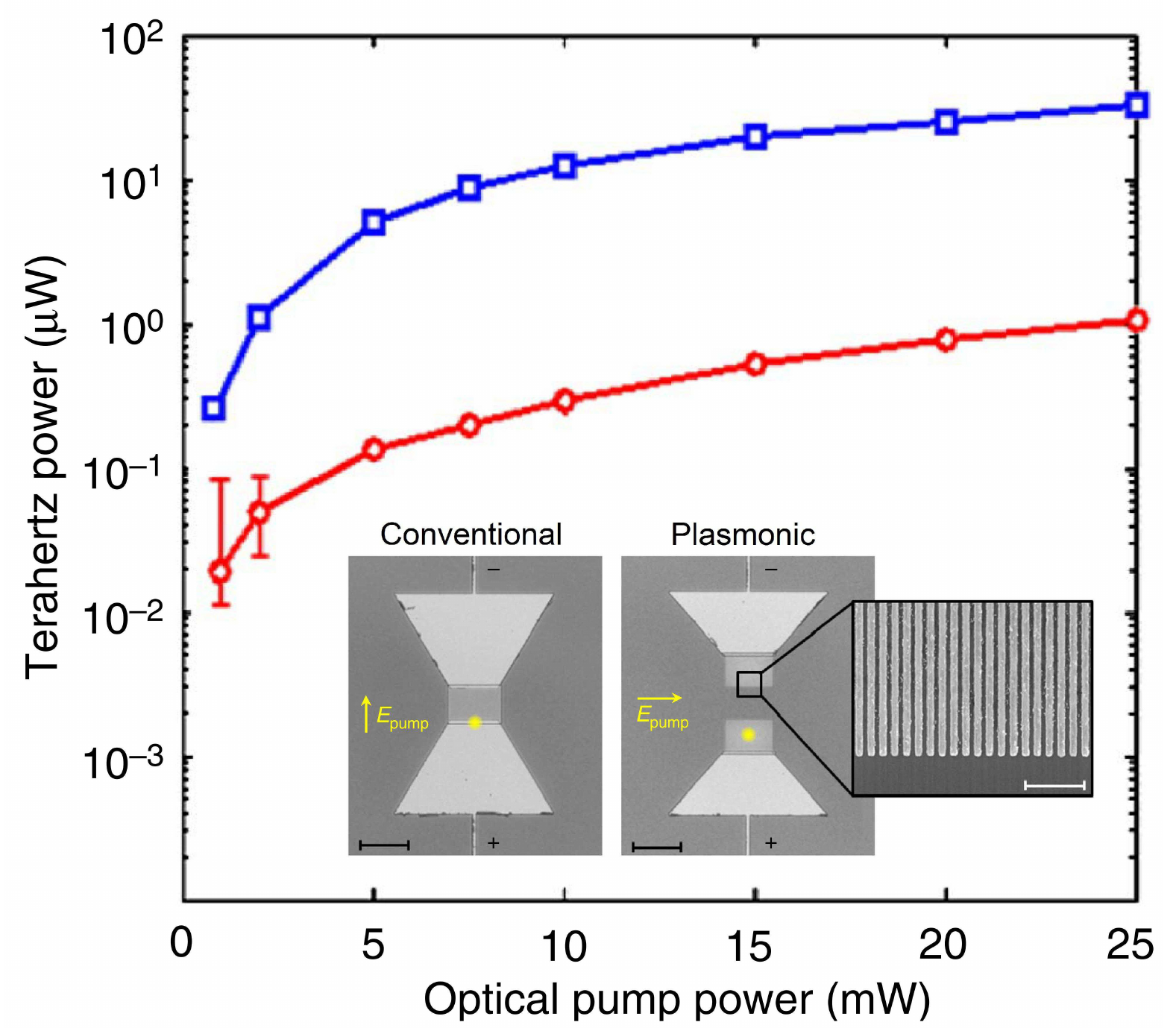}
\end{center}
\caption{THz output power vs pump power for THz hybrid optical photoconductive antenna (blue) and a standard THz photoconductive antenna of the same geometry(red).~\cite{electrodes2}}
\label{ncomm}
\end{figure}

To manufacture such structures most often methods of electron-beam lithography, accompanied by the deposition of Au on LT-GaAs substrate are used. A dielectric layer covering nanoantenna reduces Fresnel reflection from its surface, thereby increasing the absorption coefficient of the optical pump into the substrate. Usually, 200~nm thick silica ($\rm SiO_2$) is used for these purposes. This coating provides a 70\% absorption of pump light into the semiconductor material.

Comparison of traditional THz photoconductive antennae and hybrid optical-to-THz antennae with integrated plasmonic gratings reveals at least one order of magnitude superiority of the latter (Figure~\ref{ncomm}). Such a significant increase in the output power of THz radiation is achieved by reducing the path length of charge carriers to the electrodes, thus greatly increasing the photocurrent and decreasing the number of recombinated electron-hole pairs.

An important role in THz hybrid optical photoconductive antennas is played by the thickness of the semiconductor substrate, since at a sufficiently large depths (over 100~nm) photoinduced charge carriers begin to accumulate that do not participate in the effective generation of THz radiation. These electrons and holes rise from the depths of the substrate to the surface, where recombine with other charges, which significantly reduces the ratio of optical-terahertz conversion. To prevent negative effects caused by the non-equilibrium carriers accumulated in the lower layers of the substrate, its thickness should not exceed 100~nm. However, there is a workaround for this problem without reducing the thickness of semiconductor by using three-dimensional plasmonic gratings~\cite{3d75}.

\subsection{Three-dimensional plasmonic gratings} \label{second2}

\begin{figure}
\begin{center}
\includegraphics[width=0.5\textwidth]{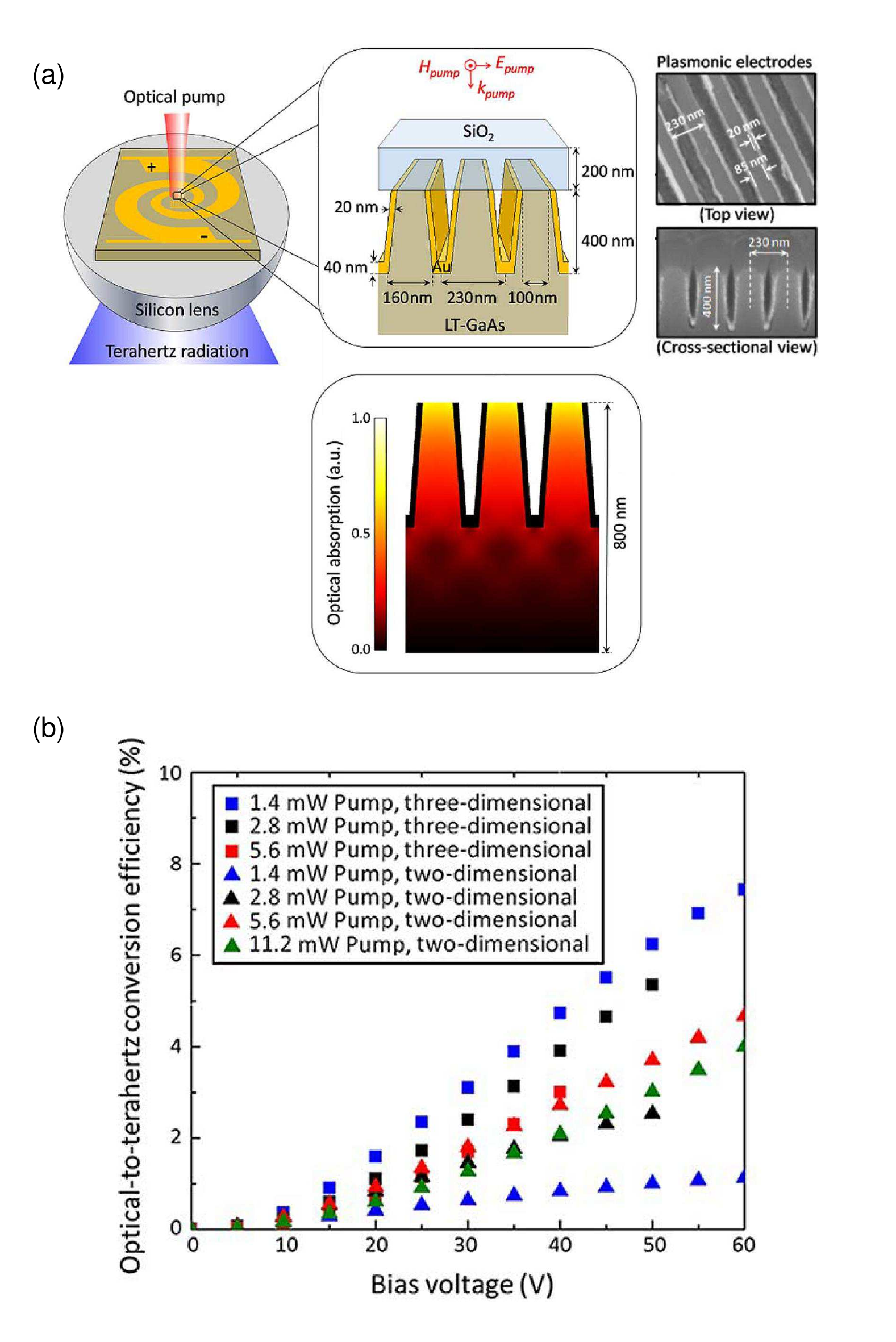}
\end{center}
\caption{(a) Schematic representation of the logarithmic hybrid THz-optical and photoconductive antenna structures with integrated three-dimensional plasmonic structure, as well as its SEM picture. (b) The dependence of the optical-terahertz conversion on bias applied to antenna electrodes at different pump powers for THz photoconductive antennas with built-in two-dimensional and three-dimensional plasmonic gratings.~\cite{3d75}}
\label{3d1}
\end{figure}

Plasmonic gratings described above are two-dimensional nano-sized structures on the semiconductor surface. They provide a high absorption of light quanta by a semiconductor material and significantly increase the concentration of photoinduced charge carriers near metal electrodes. However, the effectiveness of optical-terahertz conversion is still limited by the depth at which charge carriers are generated ($\thicksim$100~nm). To overcome this limitation, three-dimensional plasmonic gratings embedded in the semiconductor substrate are proposed. It has been shown, that such structures allow to achieve 7.5\% optical-to-THz conversion efficiency~\cite{3d75}. Figure~\ref{3d1} (a) shows a schematic of three-dimensional plasmonic gratings and distribution pattern of the absorbed optical radiation. Because of its complexity, the manufacturing of such antennas should be discussed in more detail. The process of three-dimensional plasmonic gratings fabrication starts with chemical vapor deposition of a 200~nm film of $\rm SiO_2$ onto the surface of the semiconductor. Next, on the surface of the $\rm SiO_2$ film, a nanoscale Nickel grid is applied by electron beam lithography. The grid serves as a template layer for etching of $\rm SiO_2$. When using reactive ion etching with a source of inductively coupled plasma a $\rm SiO_2$ grating is formed, which serves as a mask to create nano-walls made of semiconductor material. Plasmonic grating is obtained by deposition of gold on the sidewalls of the resulting semiconductor nanostructures. The height of the walls of the nanostructure determines the order of the modes excited in the resulting subwavelength waveguides. In such a structure waveguide TEM modes of 4th order can be excited by p-component of the femtosecond laser~\cite{3d36,3d37,3d38}. A $\rm SiO_2$ mask is then removed by wet etching. Then to reduce the losses associated with reflection, the surface of the antenna is covered with a 200~nm layer of $SiO_2$. The resulting structure absorbs up to 70 percent of the incident laser radiation~\cite{3d38}.

Figure~\ref{3d1} (b) demonstrates the advantages of three-dimensional gratings two-dimensional It is seen that three-dimensional plasmonic grating has a large coefficient of optical-to-THz conversion in comparison with the two-dimensional analogue. At the optical pump power of 1.4~mW, optical-to-THz conversion efficiency reaches 7.5\%. Such efficiency increase is due to the fact that three-dimensional plasmonic grating embedded into the surface layer of the semiconductor substrate, hence the contact surface of three-dimensional grating with the semiconductor is much greater than that of two-dimensional plasmonic gratings. This increases the concentration of the photoinduced charges near the metal electrodes, which are involved in the effective generation of THz radiation~\cite{3d75}. However, with increasing optical pump power, optical-to-THz conversion efficiency in three-dimensional structures decreases. This happens because large number of separated electron-hole pairs generate a significant electric field with direction opposite to the external field. The resulting field turns out to decrease, hence the efficiency of THz generation is reduced~\cite{3d42}.

Thus, the THz photoconductive antennas with integrated three-dimensional plasmonic gratings provide greater efficiency of optical-to-THz conversion in comparison with two-dimensional plasmonic gratings. In the future, despite the complexity of fabrication technology, these three-dimensional nanoscale structures can replace two-dimensional in hybrid THz-optical photoconductive antennas.

\section{Discussion and outlook} \label{third}

\begin{table*}[!ht]
 \begin{tabular}{| P{3.7cm} | P{0.8cm} | P{1.8cm} |P{2cm} | P{2.2cm} | P{1.8cm} | P{1.2cm} |}
 \hline
Method of Plasmon Resonance Obtaining &Regime & Fabrication Complexity &Generation Enhancement Factor &Maximum Conversion Efficiency &At Pump Power &\# \hspace{1mm} Ref. \\
\hline
Silver Nanoparticles &Pulse&Relatively Low &2x & - &50 mW &\cite{mono1} \\
\hline
AuGe Nanoparticles &Pulse&Relatively Low &4x & - &no data &\cite{monop} \\
\hline
Hexagonal Nanoantennas &Pulse&High &7x &no data &2 mW &\cite{gex1} \\
\hline
Interdigitated metallic dipole nanoantenna (six-nanofinger) & CW &Relatively Low & - &0.0002\% &26 mW &\cite{dipole11} \\
\hline
Interdigitated metallic dipole nanoantenna (five-nanofinger) & CW &Relatively Low & - &0.0006\% &50 mW &\cite{dipole12} \\
\hline
Interdigitated metallic dipole nanoantenna (eight-nanofinger) & CW &Relatively Low & - &0.007\% &90 mW &\cite{dipole13} \\
\hline
Rectangular metal dipole nanoantennas & CW&High &4x &0.0005\% &55 mW &\cite{dipole1} \\
\hline
"Bow tie" metallic dipole nanoantennas &CW&High &7x &0.0008\% &55 mW &\cite{dipole1} \\
\hline
Plasmonic gratings for impulse THz generation &Pulse&High &35x--50x &0.2\% &25 mW &\cite{electrodes2} \\
\hline
Plasmonic gratings for CW-THz generation &Pulse&High &20x &0.03\% &50 mW &\cite{geom2} \\
\hline
Plasmonic gratings with grounding electrode&Pulse& High &50x &0.005\% &85 mW &\cite{electrodes1} \\
\hline
3D plasmonic gratings &Pulse&Extremely High &1500x &7.5\% &1.4 mW &\cite{3d75} \\
\hline
\end{tabular}
\caption{Comparison of the proposed fabrication methods of hybrid photoconductive antennas for efficient THz generation}
\label{tabl}
\end{table*}

To summarize all results presented in the Review, we have gathered together studies on various approaches to creation of hybrid photoconductive antennas for THz generation, in order to facilitate a direct comparison of the proposed methods (see Table~\ref{tabl}). We can conclude that the most high-performance hybrid THz photoconductive antennas are the ones based on three-dimensional plasmonic gratings with the maximum conversion efficiency about 7.5\%, at the pump power of 1.4~mW. For comparison, hybrid THz photoconductive antennas based on two-dimensional plasmonic gratings have the maximum conversion efficiency of about 0.2\% only, at a higher pump power 25~mW.

This Review has demonstrated obvious advantages of hybrid THz photoconductive antennas and photomixers over conventional ones, especially in their conversion efficiency at low pump powers. With the development of compact ultrafast fiber~\cite{Fermann2013} and semiconductor~\cite{Rafailov2007} lasers, followed by the use of new photoconductive materials suitable for the longer-wavelength pumping, such as GaBiAs~\cite{Bertulis2006}, InGaAs~\cite{Suzuki2005}, and also containing ErAs~\cite{Krotkus2005}, and InAs~\cite{Estacio2009,Gorodetsky2016, Leyman2016} quantum dots in GaAs substrates, these can lead to the creation of highly effective, miniature THz transmitters, working at room temperatures both in pulsed and CW regimes. Despite the complexity and high cost of hybrid electrodes fabrication, the total setup cost, that will exclude the extremely expensive titanium-sapphire laser, should significantly decrease.

Moreover, despite of a number of plasmonic nanoantennas' advantages associated with their small size and strong electric field localisation, such nanoantennas have large dissipative
losses resulting in their low efficiency. To overcome such limitations, a new type of \textit{nanoantennas based on dielectric nanoparticles} (e.g. silicon) with a high index dielectric constant has been recently proposed~\cite{KrasnokAPLSup2014, evlyukhin2012demonstration, makarov2015tuning, krasnok2015towards, KrasnokOE, KrasnokNanoscale}. Such nanoantennas could be very promising for hybrid THz-optical photoconductive antennas. For example, due to low dissipative losses, they have a much higher damage threshold. The typical values of damage threshold for metallic nanostructures are: gold nanorods ($\sim$70~GW/cm$^2$ or $\sim$10~mJ/cm$^2$ at 130~fs~\cite{Zayats_NN2011}), gold G-shaped nanostructure ($\sim$ 100~GW/cm$^2$ or $\sim$3~mJ/cm$^2$ at 30~fs~\cite{Valev2012}), and gold nanocylinders ($\sim$200~GW/cm$^2$ or $\sim$20~mJ/cm$^2$ at 100~fs~\cite{Zuev_2016}). According to the known data from literature, low-loss silicon nanoparticles have significantly higher damage threshold: $\sim$400~GW/cm$^2$ or $\sim$100~mJ/cm$^2$ at 250~fs~\cite{Valentine_2015}; and $\sim$1000~GW/cm$^2$ or $\sim$100~mJ/cm$^2$ at 100~fs~\cite{makarov2015tuning}. Such considerable difference in damage thresholds for plasmonic (e.g. gold) and all-dielectric (e.g. silicon) materials originates form difference in their melting temperatures (Tm(Au)=1337~K and Tm(Si)=1687 K), enthalpies of fusion (H(Au)=12.5~kJ/mol and H(Si)=50.2~kJ/mol), and absorption coefficients ($\alpha$(Au, 600~nm) = 6∙105 cm-1 and $\alpha$(Si, 600~nm) = 5∙102~cm-1). In accordance with these basic parameters, silicon has much higher damage threshold than gold. Therefore, silicon-based ultrafast all-optical modulators are more stable than plasmonic ones upon intense laser irradiation. Therefore, beside the advantages of all-dielectric nanoantennas discussed in our manuscript, they allow for applying of very high laser intensities which is promising for an effective THz generation.

Thus, despite the large number of research works in this area and very encouraging results, consensus about the advantage of a particular structure as the absorption enhancer of the pump radiation, and the magnifier of effective generation area in semiconductor is not formed yet. This is a very young research area where more studies and experiments are needed, aimed primarily at the optimisation of the proposed methods, nanoantennas, and in addition -- search for more simple solutions to the problem of transfer of optical radiation into the largest possible nonequilibrium carrier pairs number in a semiconductor, increasing the speed of their movement through the creation of more intense local electric and magnetic fields and reduction of their lifetime by reducing their free path lengths. We are sure that after successful solution of these tasks, hybrid photoconductive antennas and photomixers will bypass the efficiency of now commercially available micro-sized photoconductive antennas and photomixers so that the question of their fabrication cost will not even questioned, and they will be the one and only solution and completely take the market of compact sources of THz radiation for spectral region between 0.5 to 2.5~THz.

\section*{Conclusion}

We have discussed an approach to solving the problem of low energy conversion in THz photoconductive antennas, which comprises the use of optical nanoantennas for enhancing the efficiency of pump laser radiation absorption in the antenna gap, reducing the lifetime of photoexcited carriers and improving antenna thermal efficiency. We have described the general principles of THz generation in semiconductor sources pumped with femtosecond optical pulses and double-wavelength CW laser radiation. We have especially focused on the hybrid THz-optical photoconductive antennas and their general principles of operation. We have intended to systematise all the results obtained by researchers in this promising field of hybrid optical-to-terahertz photoconductive antennas and photomixers. We have considered the hybrid THz-optical photoconductive antennas based on plasmon monopoles, metal dipole nanoantennas, 2D and 3D plasmon gratings.

\section*{Acknowledgments}

This work was supported by the Ministry of Education and Science of the Russian Federation (Projects GZ~2014/190 and GZ~No.~3.561.2014/K), and by Russian Foundation for Basic Research (Project~16-07-01166~A). The authors are thankful to Prof.~Yuri~Kivshar, Dr.~Ilya~Shadrivov, and Mr.~Mingzhao~Song for useful discussions.


%

\end{document}